\documentclass[a4paper,11pt,fleqn,final]{article}
\pdfoutput=1 
\usepackage{jheppub} 
\usepackage[T1]{fontenc} 
\usepackage[utf8]{inputenc}
\usepackage{booktabs}
\usepackage{mathtools} 
\usepackage{xifthen}
\usepackage{siunitx} 
\usepackage{varioref}

\usepackage[noabbrev]{cleveref}
\crefname{equation}{eq.}{eqs.}%
\Crefname{equation}{Equation}{Equations}%
\usepackage{enumitem}
\setlist{noitemsep}
\setlist[itemize]{noitemsep, topsep=0pt}
\usepackage{slashed}
\usepackage{multirow}
\usepackage{myMacros}
\usepackage{lmodern}
\usepackage{subcaption}
\usepackage[olditem,oldenum]{paralist}
\newcommand{\pmns}{\text{\sc pmns}}
\newcommand{\Vpmns}{V^{\text{\sc pmns}}}

\title{Reinterpreting the ATLAS bounds on heavy neutral leptons in a realistic neutrino oscillation model}

\author[a,b]{J.-L.\ Tastet}
\author[a]{O.\ Ruchayskiy}
\author[b]{I.\ Timiryasov}

\affiliation[a]{Niels Bohr Institute, University of Copenhagen,\\
  Blegdamsvej 17, DK-2010, Copenhagen, Denmark}
\affiliation[b]{Institute of Physics, Laboratory for Particle Physics and Cosmology,\\
  École Polytechnique Fédérale de Lausanne, CH-1015 Lausanne, Switzerland}

\emailAdd{jean-loup.tastet@epfl.ch}
\emailAdd{Oleg.Ruchayskiy@nbi.ku.dk}
\emailAdd{Inar.Timiryasov@epfl.ch}


\arxivnumber{2107.12980}

\abstract{%
Heavy neutral leptons (HNLs) are hypothetical particles, motivated in the
first place by their ability to explain neutrino oscillations. 
Experimental searches for HNLs are typically conducted under the assumption of a
single HNL mixing with a single neutrino flavor.
However, the resulting exclusion limits may not directly constrain the corresponding mixing angles in \emph{realistic HNL models} --- those which can explain neutrino oscillations.
The reinterpretation of the results of these experimental searches turns out to be a non-trivial task, that requires significant knowledge of the details of the experiment.
In this work, we perform a reinterpretation of the latest ATLAS search for HNLs decaying promptly to a tri-lepton final state.
We show that in a realistic model with two HNLs, the actual limits can vary by several orders of magnitude depending on the free parameters of the model.
Marginalizing over the unknown model parameters leads to an exclusion limit on the total mixing angle which can be up to 3 orders of magnitude weaker than the limits reported in ref.~\cite{Aad:2019kiz}.
This demonstrates that the reinterpretation of results from experimental searches is a \emph{necessary} step to obtain meaningful limits on realistic models.
We detail a few steps that can be taken by experimental collaborations in order to simplify the reuse of their results.
}

\begin{document}

\maketitle
\flushbottom

\providecommand{\nuMSM}{\ensuremath{\mathrm{\nu MSM}}}
\providecommand{\nufit}{NuFIT~5.0}
\providecommand{\CLs}{\ensuremath{\mathrm{CL}_s}}
\providecommand{\pT}{\ensuremath{p_{\mathrm{T}}}}


\section{Introduction}
\label{sec:intro}

\subsection{Heavy neutral leptons}

The idea that new particles need not be heavier than the electroweak scale, but rather can be light and feebly interacting is drawing increasing attention from both the theoretical and experimental communities~\cite[see e.g.\@][]{Shaposhnikov:2007nj,Alekhin:2015byh,Beacham:2019nyx,Strategy:2019vxc}.
In particular, the hypothesis that heavy neutral leptons are responsible for (some of the) beyond-the-Standard-Model phenomena has been actively explored in recent years, see e.g.\ \cite{Asaka:2005an,Asaka:2005pn,Shaposhnikov:2007nj,Boyarsky:2009ix,Drewes:2013gca,Alekhin:2015byh,Deppisch:2015qwa} and refs.\ therein. 
Heavy neutral leptons (HNLs) are massive particles that interact similarly to neutrinos, but with their interaction strength suppressed by flavor-dependent dimensionless numbers --- \emph{mixing angles} --- ($U_e^2, U_\mu^2, U_\tau^2$).
HNLs
first appeared in the context of left-right symmetric models
  \cite{Pati:1974yy,Mohapatra:1974gc,Mohapatra:1974hk,Senjanovic:1975rk} which required an extension of the fermion sector with Standard Model (SM) gauge singlet particles, and then in the (type I) see-saw mechanism
  \cite{Minkowski:1977sc,Yanagida:1979as,Glashow:1979nm,GellMann:1980vs,Mohapatra:1979ia,Mohapatra:1980yp,Schechter:1980gr,Schechter:1981cv} in which heavy Majorana neutrinos lead to light Standard Model neutrinos.
The interest for these models increased when it was recognized that the same particles could also be responsible for the generation of the matter-antimatter asymmetry of the Universe~\cite{Fukugita:2002hu}.
This scenario (known as \emph{leptogenesis}) has been actively developed since the 1980s (see reviews~\cite{Davidson:2008bu,Shaposhnikov:2009zzb}).
In particular, it was found that the Majorana mass scale of right-handed neutrinos could be as low as the $\si{TeV}$, $\si{GeV}$ or even $\si{MeV}$ scale~\cite{Akhmedov:1998qx,Pilaftsis:2009pk,Asaka:2005pn,Shaposhnikov:2008pf,Canetti:2010aw,Drewes:2017zyw}; for a recent overview see~\eg{}~\cite[]{Klaric:2020lov, Klaric:2021cpi}.
While two HNLs are sufficient to explain neutrino masses and oscillations as well as the origin of the matter-antimatter asymmetry, a third particle can play the role of dark matter~\cite{Asaka:2005an,Asaka:2005pn,Boyarsky:2009ix,Boyarsky:2018tvu,Ghiglieri:2020ulj} within the \emph{Neutrino Minimal Standard Model} ($\nu$MSM).

Starting from the 1980s~\cite{Shrock:1980vy,Shrock:1980ct,Shrock:1981wq,Gronau:1984ct}, many experiments have searched for HNLs (as summarized \eg{} in refs.~\cite{Gorbunov:2007ak,Atre:2009rg,Alekhin:2015byh,Deppisch:2015qwa,Bryman:2019bjg,Beacham:2019nyx}).
Current generation particle physics experiments, including LHCb, CMS, ATLAS, T2K, Belle and NA62, all include HNL searches into their scientific programs~\cite{Liventsev:2013zz,Aaij:2014aba,Artamonov:2014urb,Aad:2015xaa,Khachatryan:2015gha,Gligorov:2017nwh,CortinaGil:2017mqf,Mermod:2017ceo,Izmaylov:2017lkv,Sirunyan:2018mtv,Aad:2019kiz,NA62:2020mcv,CortinaGil:2021gga}. 
However, as pointed out in ref.~\cite{Abada:2018sfh}, most of the existing or proposed analyses concentrate on the case of a \emph{single HNL mixing with only one flavor}. 
Such a model serves as a convenient benchmark, but it cannot explain any of the BSM phenomena that served as initial motivations for postulating HNLs.
The same benchmarks are used when estimating the sensitivity of future experiments \cite[see \eg{}][]{Beacham:2019nyx}, with the notable exception of the SHiP experiment, which provided sensitivity estimates for arbitrary sets of mixing angles~\cite{SHiP:2018xqw}.
This raises a few questions:
\begin{enumerate}
    \item Which HNL models explaining neutrino oscillations and/or other BSM phenomena are allowed or ruled out by previous searches? What parts of the HNL parameter space will be probed by future experiments?
    \item What information do experimental groups need to provide in order to facilitate the answer to such questions in the future?
\end{enumerate}
A number of tools exists, see \eg{} \cite{Cranmer:2010hk,Buckley:2010ar,Conte:2012fm,Conte:2013mea,Drees:2013wra,Kraml:2013mwa,Papucci:2014rja,Barducci:2014ila}, that allow recasting LHC results for new sets of models (see also~\cite{Abdallah:2020pec}).
These tools have mostly been developed in the context of supersymmetry and similar searches at the LHC and are not readily applicable to HNL models, whose collider phenomenology is quite different.

\textbf{In this work} we perform a step in the direction of \emph{recasting LHC results}.
Specifically, we recast the ATLAS tri-lepton search~\cite{Aad:2019kiz} in the case of the \emph{simplest realistic HNL model of neutrino oscillations}.
This model features \textit{two heavy neutral leptons} with (almost) degenerate masses.
The possible values of the HNL mixings are constrained by neutrino oscillation data.%
\footnote{In the case of three or more HNLs, the constraints on the HNL mixing angles are much more relaxed thanks to the freedom conferred by the additional model parameters~\cite{Abada:2018oly,Chrzaszcz:2019inj}. Such models can thus accommodate more extreme ratios of the mixing angles (and, for four or more HNLs, even allow some mixing angles to be zero). However, many of the results that we will discuss in this paper still apply. In particular, most points in the parameter space of these models also correspond to non-trivial mixing patterns, and in order to probe them the ATLAS results will need to be recast.}
In what follows we will refer to this model as a \emph{realistic HNL model}.
As we shall see below, even in this simple model, the interpretation of the results is a non-trivial task.

\subsection{Motivation for a reinterpretation}
\label{sec:motivation}

\begin{figure}
    \centering
    \begin{subfigure}[b]{0.48\textwidth}
        \centering
        \includegraphics[width=0.85\textwidth]{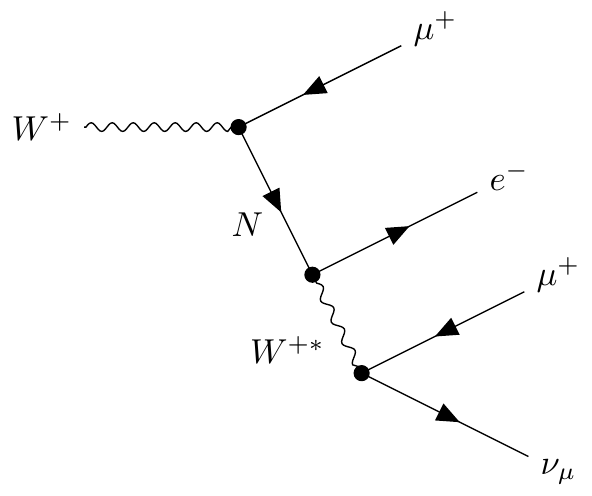}\par
        \caption{LNC}
        \label{fig:lnc_process}
    \end{subfigure}
    \begin{subfigure}[b]{0.48\textwidth}
        \centering
        \includegraphics[width=0.85\textwidth]{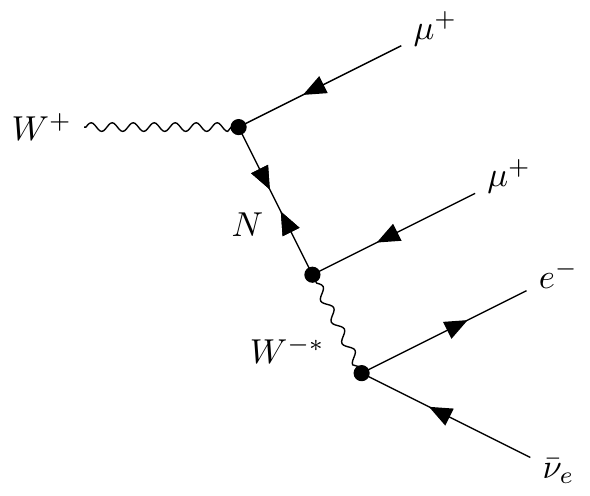}\par
        \caption{LNV}
        \label{fig:lnv_process}
    \end{subfigure}
    \caption{Lepton number conserving (LNC) and violating (LNV) diagrams contributing to the same $\mu^+ \mu^+ e^-$ + missing transverse energy (MET) final state.}
    \label{fig:LNV_LNC}
\end{figure}

The realistic seesaw model describing neutrino oscillations brings several changes compared to the single-HNL, single-flavor model analyzed by the ATLAS collaboration~\cite{Aad:2019kiz}.
The analysis from ref.~\cite{Aad:2019kiz} concentrated on the following process:
\begin{equation}
    \label{eq:trilepton}
    pp \to W^\pm +X\quad\text{with}\quad W^\pm \to \ell_\alpha^\pm + N \quad\text{followed by}\quad N \to \ell_\alpha^\pm +\ell_\beta^\mp +\smash{\overset{\scriptscriptstyle(-)}{\nu}_{\!\!\beta}}
\end{equation}
where $\ell^\pm_\alpha$ are light leptons ($e^\pm$ or $\mu^\pm$), $\alpha \neq \beta$ and \smash{$\overset{\scriptscriptstyle(-)}{\nu}_{\!\!\beta}$} is a neutrino or anti-neutrino with flavor $\beta$.
They performed two independent analyses: one for the $e^\pm e^\pm \mu^\mp$+MET final state (``\emph{electron channel}'') and one for the $\mu^\pm \mu^\pm e^\mp$+MET final state (``\emph{muon channel}'').
In both cases, only a single process (corresponding to diagram (b) in \cref{fig:LNV_LNC}), along with its CP-conjugate, contributed to the final signal.
The upper limit on an admissible signal was thus directly translated into an upper bound on the mixing angle $U_e^2$ or $U_\mu^2$, depending on the channel.
The situation changes once we consider a realistic seesaw model with 2 HNLs:
\begin{enumerate}
    \item In such a model, several processes contribute incoherently\footnote{Their diagrams all produce different final states (when taking the light neutrino and its helicity into account) and therefore they do not interfere.} to each final state.
    The upper bound on an admissible signal in any channel thus translates non-trivially into limits on all three mixings angles ($U_e^2, U_\mu^2, U_\tau^2$). 
    
    \item Any set of mixing angles consistent with neutrino oscillation data leads to observable signals in both the $e^\pm e^\pm \mu^\mp$ and $\mu^\pm \mu^\pm e^\mp$ channels, therefore the statistical procedure should take into account that the signal is non-zero in both channels.
    
    \item Different processes that contribute to the same tri-lepton final state have different kinematics (due in part to spin correlations~\cite{Tastet:2019nqj}). Therefore the signal efficiencies need to be evaluated separately for every process.
    
    \item We consider 2 HNLs with \emph{nearly degenerate} masses. 
    Due to HNL oscillations~(\cf{} \cite{Tastet:2019nqj} or \cite{Boyanovsky:2014una,Cvetic:2015ura,Anamiati:2016uxp,Antusch:2017ebe,Das:2017hmg,Cvetic:2018elt,Hernandez:2018cgc} for earlier works) tiny mass differences (well below the mass resolution limit of ATLAS) can significantly affect the interference pattern, leading to the suppression or enhancement of some processes as compared to the single HNL case, see \eg{}~\cite{Shaposhnikov:2006nn,Kersten:2007vk,Anamiati:2016uxp,Drewes:2019byd}.
    Since different processes (such as those in \cref{fig:LNV_LNC}) have different kinematics and thus efficiencies, this implies that the overall signal efficiency depends not only on the mixing angles, but also on the level of the HNL mass degeneracy.
    In order to account for this, we present our analysis for two limiting cases: the ``Majorana-like'' and ``Dirac-like'' limits (which we will define in \cref{sec:seesaw}).
\end{enumerate}
All these points make it impossible to reinterpret the ATLAS results by just rescaling them (as done \eg{} in ref.\ \cite{Bondarenko:2021cpc}).
Instead one should perform a full signal and background modeling and evaluate the signal selection efficiencies.
Although this can only be done properly by the collaboration itself, thanks to their access to the full detector simulation, the analysis framework and the actual counts in the signal regions, we will demonstrate that one can nonetheless reproduce the original ATLAS limits sufficiently well for the purpose of reinterpretation.
Finally, we will briefly discuss what data from the collaboration could simplify our analysis and make it more precise, in the spirit of the recommendations from the LHC Reinterpretation Forum~\cite{Abdallah:2020pec}.

\paragraph{The present paper is organized as follows:}
In \cref{sec:seesaw} we introduce the notion of ``realistic'' seesaw models. To this end, we review the so-called type-I seesaw mechanism, discuss how neutrino oscillation data constrain its parameters, and examine how interference effects between multiple HNLs can completely change their phenomenology.
We then describe our analysis procedure in \cref{sec:procedure}: we present the event selection, detail the calculation of the expected signal and efficiencies, and discuss our background model as well as the statistical method used to derive the exclusion limits.
In \cref{sec:results}, we finally present our reinterpretation of the ATLAS limits on promptly-decaying HNLs within a realistic seesaw model with 2 HNLs, and we comment on these results.
We conclude in \cref{sec:conclusion}, and summarize what data should ideally be reported by experiments in order to allow reinterpreting their limits easily and accurately within realistic models.

\section{Realistic neutrino oscillation models}
\label{sec:seesaw}

\subsection{The Lagrangian of the model}

Our starting point is the type I seesaw mechanism~\cite{Minkowski:1977sc,Yanagida:1979as,GellMann:1980vs,Mohapatra:1979ia,Schechter:1980gr,Schechter:1981cv}, that we briefly review below.
The exposition is fairly standard and can be found, \eg{} in refs.~\cite{Boyarsky:2009ix,Alekhin:2015byh,Boyarsky:2018tvu} and \cite[][ch.~14]{Zyla:2020zbs}.
The reader can skip it, taking notice of the definitions \eqref{mixing}--\eqref{eq:theta}.

The Lagrangian of the model reads
\begin{equation}
	\mathcal{L}_{\mathrm{SM}+\mathrm{HNL}}= \mathcal{L}_\mathrm{SM}
	+ i \bar{\nu}_{R_I} \slashed{\partial} \nu_{R_I}
	- F_{\alpha I} (\bar{L}_\alpha \cdot \tilde{\Phi}) \nu_{R_I}
	- \frac{1}{2} M_I \bar{\nu}^c_{R_I}\nu_{R_I},
	\label{eq:lagrangian}
\end{equation}
where $\mathcal{L}_{\mathrm{SM}}$ is the usual SM Lagrangian and $\nu_{R_I}$ are new right-handed particles that are SM gauge singlets. In the present paper we will consider the case of two HNLs, therefore the index $I$ runs over $1$, $2$.
$L_\alpha$ are the left-handed lepton doublets labeled with
the flavor index $\alpha = e, \mu, \tau$ and $\tilde\Phi = i\sigma_2 \Phi$,
where $\Phi$ is the Higgs doublet. $F_{\alpha I}$ is the matrix of
Yukawa couplings in the basis where the Yukawa couplings of charged leptons and the Majorana mass $M_I$
of the right-handed neutrinos are both diagonal.
After electroweak symmetry breaking, the Higgs field in the Lagrangian~\eqref{eq:lagrangian} obtains a vacuum expectation value $\langle \Phi \rangle = (0,\,v)^T$ and
the Yukawa interaction terms in~\cref{eq:lagrangian} effectively become Dirac mass terms coupling the left and right chiral components of the neutrinos.
Since the right-handed neutrinos have, in addition, a Majorana mass, the spectrum of the theory is obtained by diagonalizing the full mass matrix.

For $|F_{\alpha I} v|\ll |M_I|$ one finds after the diagonalization 3 light mass eigenstates $\nu_i$ with masses ${m_1,m_2,m_3}$ and two 
heavy mass eigenstates $N_I$ --- the HNLs --- with masses $M_1$ and $M_2$.\footnote{Given the Lagrangian~\eqref{eq:lagrangian} with two right-handed neutrinos, the lightest neutrino is massless (up to quantum corrections~\cite{Davidson:2006tg}).}
As a consequence, the flavor eigenstates (SM neutrinos) $\nu_{L \alpha}$ can be expressed as a linear combination of the 5 mass eigenstates as
\begin{equation}
	\nu_{L \alpha} = 
	V^{\pmns}_{\alpha i}\nu_i + \Theta_{\alpha I} N_I^c\,,
	\label{mixing}
\end{equation}
where $\Vpmns$ is the 
Pontecorvo-Maki-Nakagawa-Sakata (PMNS) matrix~(see \eg{} \cite{bilenky:2014ema}).
As a result, the heavy mass eigenstates $N_I$ contain an admixture of SM neutrinos $\nu_{L \alpha}$, and therefore possess ``weak-like'' interactions, suppressed
by the mixing angles $\Theta_{\alpha I}$, approximately given by
\begin{equation}
    \label{eq:theta}
	\Theta_{\alpha I} \simeq \frac{v F_{\alpha I}}{M_I}\,.
\end{equation}

\subsection{Parametrization of the Yukawas}
\label{sec:mixings}

The Lagrangian~\eqref{eq:lagrangian} contains 11 new parameters, as compared to the SM one~\cite{Alekhin:2015byh}.
These parameters are, however, constrained by neutrino oscillation data~\cite{Esteban:2018azc}.
Five neutrino parameters have already been measured: two mass differences ($\Delta m_{\rm atm}^2$ and $\Delta m_{\rm sun}^2$) and three mixing angles ($\theta_{12},\theta_{23},\theta_{13}$).
The remaining unknown parameters are the mass of the lightest neutrino, two Majorana phases, and the $CP$-violating phase $\delta$. Our \textit{a priori} choice of two HNLs restricts the mass of the lightest neutrino to be zero and only allows a certain combination of the Majorana phases to be independent. 
As a result, we are left with only two unknown parameters in the active neutrino sector, in addition to the discrete choice of the mass ordering.\footnote{%
	These parameters may be probed in a not-so-distant future: for the inverted hierarchy, the next generation of neutrinoless double beta decay experiments may provide information on the Majorana phases~\cite{Giuliani:2019uno}, while the $CP$-violating phase $\delta$ is already constrained by T2K~\cite{Abe:2019vii}, with further improvements expected from the DUNE experiment~\cite{Abi:2018dnh}.
}

The measured low-energy parameters mean that for any choice of heavy neutrino masses $M_I$, the Yukawa couplings $F_{\alpha I}$ are not completely free.
To account for this, we can parametrize the neutrino Yukawa couplings using the Casas-Ibarra parametrization~\cite{Casas:2001sr}:
\begin{align}
	F = \frac{i}{v} \Vpmns \sqrt{\strut m_\nu^\mathrm{diag}} R \sqrt{\strut M^\mathrm{diag}}\,,
	\label{parametrization:CI}
\end{align}
where the matrix $M^\mathrm{diag}=\diag{M_1,M_2}$, and $R$ is a complex $3\times 2$ matrix satisfying $R^{\mathrm{T}} R=1_{2\times 2}$.
For the PMNS matrix we use the standard parametrization~\cite{Zyla:2020zbs}.
We parametrize the relevant combination of the Majorana phases in the PMNS matrix as $\eta = \frac12 (\alpha_{21} - \alpha_{31} )$ for the normal neutrino mass hierarchy (NH), and $\eta = \frac12 \alpha_{21} $ for the inverted hierarchy (IH), with $\eta \in [0,2\pi[$.
The light neutrino mass matrix is $m_\nu^\mathrm{diag}=\mathrm{diag}(m_1,m_2,m_3)$ with $m_1=0$ for NH, and $m_3=0$ for IH.

In the model with two right-handed neutrinos, the matrices $R$ depend on the neutrino mass hierarchy and are given by
\begin{align}
	R^{\rm NH}=
	\begin{pmatrix}
		0 && 0\\
		\cos \omega && \sin \omega \\
		-\xi \sin \omega && \xi \cos \omega
	\end{pmatrix}\,,\quad \quad
	R^{\rm IH}=
	\begin{pmatrix}
		\cos \omega && \sin \omega \\
		-\xi \sin \omega && \xi \cos \omega \\
		0 && 0
	\end{pmatrix}
	\,.
\end{align}
with a complex angle $\omega=\Re\omega+i \Im\omega$, and a discrete parameter $\xi=\pm1$.
Changing the sign of $\xi$ can be undone by $\omega \to - \omega$ along with $ N_2 \to - N_1$~\cite{Abada:2006ea}, so we fix $\xi = +1$.

\subsection{Heavy neutrino mixing}

The weak-like interactions of HNLs are suppressed by the \textit{mixing angles} $\Theta_{\alpha I}$ defined in \cref{eq:theta}.
These mixing angles may contain complex phases, which play no role for the processes that we consider.\footnote{These complex phases can be important if \textit{the period of HNL oscillations} is comparable with the size of the experiment, see \eg{}~\cite{Tastet:2019nqj} and references therein.}
Only the cumulative effects of both $N_1$ and $N_2$ contributes to the observed signal and therefore the experimentally measurable quantities are
\begin{align}
	U_\alpha^2 \equiv \sum_I |\Theta_{\alpha I}|^2 \quad \text{ and }\quad
	U_{\text{tot}}^2 \equiv \sum_{\alpha, I} |\Theta_{\alpha I}|^2\,,
	\label{U2:def}
\end{align}
which respectively quantify the total HNL mixing to a particular flavor and the overall mixing between HNLs and neutrinos of definite flavor.
The latter quantity has a particularly simple form in terms of the neutrino masses and Casas-Ibarra parameters:
\begin{align}
	U_{\text{tot}}^2 = \frac{\sum_i m_i}{M_N} \cosh \left( 2 \Im \omega \right)
	\label{eq:U2Imomega}
\end{align}
where $M_N = \frac 12(M_1 + M_2)$.
For the corresponding expressions of $U_\alpha^2$, see \eg{} ref.~\cite{eijima:2018qke}.

\begin{figure}
    \centering
    \includegraphics[width=0.85\textwidth]{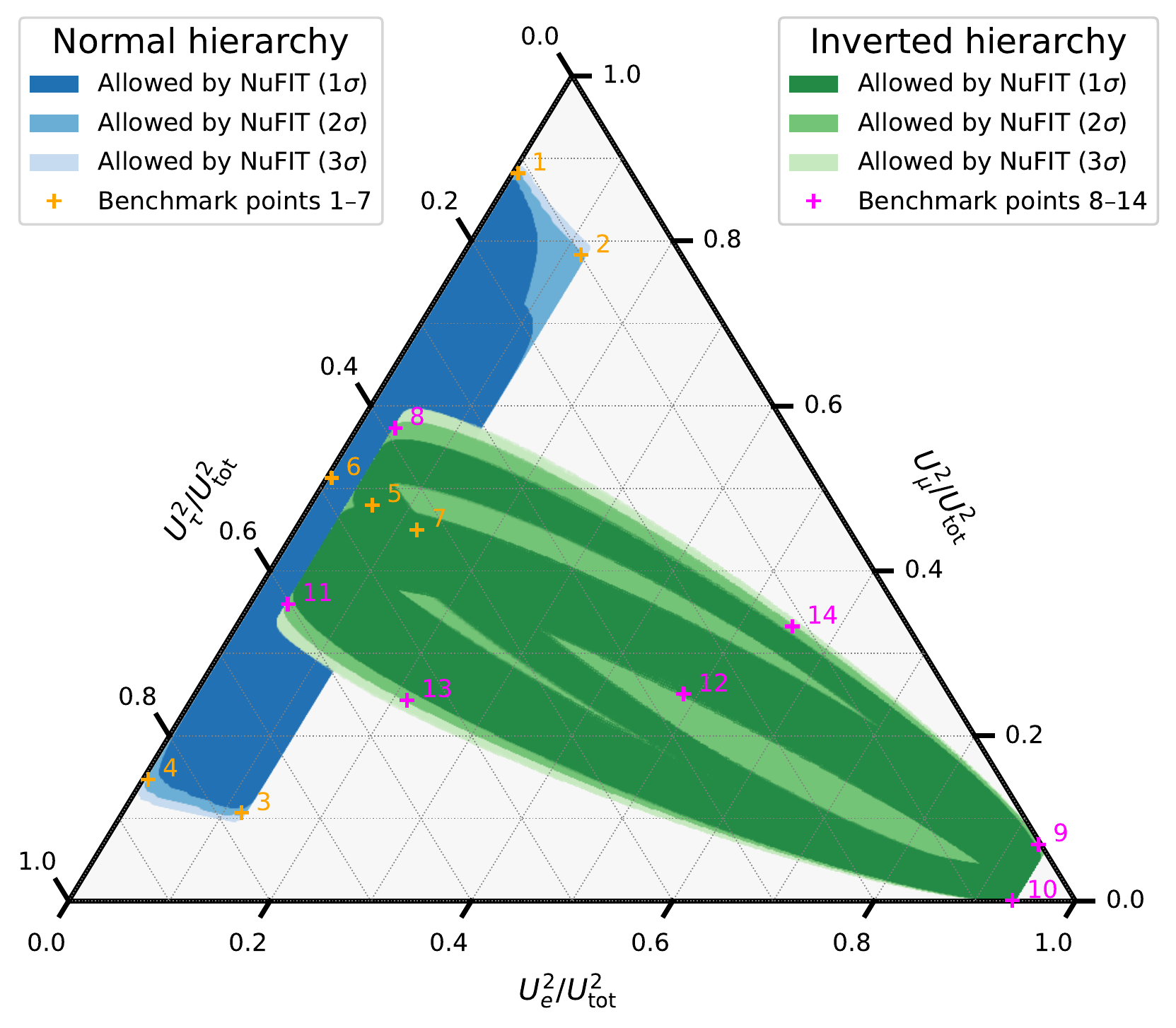}
    \caption{Ternary plot showing the combinations of mixing angles $U_{\alpha}^2/U_{\text{tot}}^2$, $\alpha=e,\mu,\tau$, which are consistent with the \nufit{} \cite{Esteban:2020cvm,NuFIT5.0} fit to neutrino oscillation data, at the $1$, $2$ and $3\sigma$ levels, for the normal and inverted hierarchies. The markers denote the selected benchmark points, which are meant to represent both typical and extreme ratios of the squared mixing angles.}
    \label{fig:ternary_plot_with_benchmarks}
\end{figure}

As we have already mentioned, not all values of the Yukawa couplings $F_{\alpha I}$ --- and hence of $U_\alpha^2$ --- are compatible with neutrino oscillation data. Only certain regions are allowed in $(U_e^2,\, U_\mu^2,\, U_\tau^2)$ space. For $|\Im \omega| \gg 1$ and $|M_1 - M_2| \ll M_N$, the shape of these regions \textit{does not depend}
on $M_N$, $|M_1 - M_2|,$ or $U^2_{\text{tot}}$.
Taking into account that
\begin{equation}
    U_{e}^2/U_{\text{tot}}^2 + U_{\mu}^2/U_{\text{tot}}^2 + 
    U_{\tau}^2/U_{\text{tot}}^2 = 1 \,,
\end{equation}
we can display the combinations of $U_\alpha^2$ which are compatible with neutrino oscillation data using a ternary plot as in \cref{fig:ternary_plot_with_benchmarks}, \cf{}~\cite{Drewes:2016jae,Caputo:2017pit,Drewes:2018gkc,Bondarenko:2021cpc}.
In our analysis, we used the most recent global fit to neutrino oscillation data, \nufit{} \cite{Esteban:2020cvm,NuFIT5.0}.
The shape of the allowed regions depends on the values of the Dirac phase $\delta$ and of the active neutrino mixing angle $\theta_{23}$. We have used the three-dimensional projections of $\Delta \chi^2$ provided by \nufit{} in order to determine the $1$, $2$ and $3\sigma$ contours presented in \cref{fig:ternary_plot_with_benchmarks}.\footnote{We have used the $\Delta\chi^2$ tables from \nufit{} which do not include the atmospheric data provided by the Super-Kamiokande collaboration~\cite{Abe:2017aap}.
Our choice of benchmark models is only slightly affected by this choice, and this does not qualitatively change our analysis or conclusions.}
In order to better visualize the correspondence between the exclusion limits and various points in the allowed regions, we have defined a number of benchmarks, which are represented in \cref{fig:ternary_plot_with_benchmarks}.

\subsection{Quasi-Dirac HNLs, lepton number violating effects and relevant limits}
\label{sec:quasi_dirac_limit}

As neutrino oscillations do not constrain the masses of HNLs, $M_1$ and $M_2$ can be arbitrary.
In this work we choose to consider the case where $M_1 \approx M_2$, \ie{}
\begin{equation}
    \label{eq:DeltaM}
    \Delta M \equiv |M_1 - M_2| \ll M_N = \frac{M_1+M_2}{2}.
\end{equation}
The motivation for this scenario is twofold. First, the mass degeneracy of two HNLs allows for sizable mixings between active neutrinos and HNLs in a technically natural way~\cite{Wyler:1982dd,Mohapatra:1986bd,Branco:1988ex,GonzalezGarcia:1988rw,Shaposhnikov:2006nn,Kersten:2007vk,Abada:2007ux,Roy:2010xq,Gavela:2009cd,Drewes:2019byd}. Secondly, low-scale leptogenesis (see the recent work~\cite{Klaric:2020lov} and references therein) requires a mass degeneracy between two heavy neutrinos. The mass splitting between the HNLs needs to be especially tiny if one wants to create the initial conditions required for the generation of sterile neutrino dark matter in the early Universe~\cite{Shaposhnikov:2008pf,Canetti:2012kh,Ghiglieri:2020ulj}.

In the limit $M_1 \approx M_2$ there is an approximate global $U(1)$ symmetry in the theory.\footnote{The symmetry becomes exact when $M_1 = M_2$ and $\Theta_{\alpha 1} = \pm i \Theta_{\alpha 2}$. In this limit active neutrinos become massless and the two HNLs form a single Dirac particle $\Psi$ such that $\frac{1+\gamma_5}2 \Psi = \frac{\nu_{R_1} + i \nu_{R_2}}{\sqrt{2}}$.}
In this \textit{quasi-Dirac limit} of the two-HNLs model, the lepton number violating (\textbf{LNV}) processes (such as \ref{fig:LNV_LNC}(b)) are suppressed compared to the lepton number conserving (\textbf{LNC}) processes.
When $M_1\neq M_2$ but $\Delta M \ll M_N$, HNL oscillations take place, as discussed in \eg{}~\cite{Asaka:2005pn,Boyanovsky:2014una,Cvetic:2015ura,Anamiati:2016uxp,Antusch:2017ebe,Das:2017hmg,Cvetic:2018elt,Hernandez:2018cgc,Tastet:2019nqj}.
As a result, lepton number violation may not be suppressed any more. Rather, the rates of LNC and LNV processes undergo a periodic modulation as a function of the proper time $\tau = \sqrt{(x_{\mathrm{D}}-x_{\mathrm{P}})^2}$ between the HNL production and decay vertices~\cite{Tastet:2019nqj}:
\begin{equation}
  \diff\Gamma_{\alpha\beta}^{\text{\tiny\sc lnc/lnv}}(\tau) \cong
  2 \abs{\Theta_{\alpha 1}}^2 \abs{\Theta_{\beta 1}}^2 \Bigl( 1 \pm \cos\left(\Delta M \tau\right) \Bigr) e^{-\Gamma \tau}
  \diff\hat{\Gamma}_{\alpha\beta}^{\text{\tiny\sc lnc/lnv}}
  \label{eq:quasi_dirac}
\end{equation}
with the ($+$) sign for LNC and ($-$) for LNV, and where $\diff\hat{\Gamma}_{\alpha\beta}^{\text{\tiny\sc lnv/lnc}}$ is the differential rate for a tri-lepton process mediated by a single Majorana HNL~$N$ in the (unphysical) limit of a unit mixing angle between the HNL and the active flavor $\alpha$ at its production vertex, with flavor $\beta$ at its decay vertex, and without the absorptive part; where $\Gamma \eqdef \Gamma_1 \cong \Gamma_2$ and by assumption $\Theta_{\alpha 2} \cong \pm i\Theta_{\alpha 1}$.
Notice how in this quasi-Dirac limit, the oscillation pattern does not explicitly depend
on the lepton flavors $\alpha$ and $\beta$, but only on whether the process is LNC or LNV.
If $\Delta M$ vanishes exactly, then HNLs form a Dirac fermion and LNV effects are completely absent.
Equation~\eqref{eq:quasi_dirac} demonstrates the two limiting cases of the two-HNLs seesaw model:
\begin{equation}
    \label{eq:limits}
    \begin{aligned}
    &\Delta M \tau \ll 2\pi \quad \text{(Dirac-like limit)} &\diff\Gamma_{\alpha\beta}^{\text{\tiny\sc lnv}} \approx 0,\ \diff\Gamma_{\alpha\beta}^{\text{\tiny\sc lnc}} \text{ is enhanced by } \cong 2\\
    &\Delta M \tau \gg 2\pi \quad \text{(Majorana-like limit)} & \text{integrated partial widths } \Gamma_{\alpha\beta}^{\text{\tiny\sc lnv}}\cong \Gamma_{\alpha\beta}^{\text{\tiny\sc lnc}}
    \end{aligned}
\end{equation}
where $\tau$ must satisfy both $\tau \Gamma \lesssim 1$ and $\gamma\tau \lesssim L_{\text{det}}$ (whichever is stronger), with $\Gamma$ denoting the total HNL width, $\gamma$ its boost factor, and $L_{\text{det}}$ the typical detector size.

\bigskip\noindent\textbf{In this work} we will consider these two limiting cases for quasi-Dirac HNLs:
\begin{itemize}
    \item \textbf{Dirac-like}: the pure Dirac ($\Delta M = 0$) limit where all LNV effects are completely absent, and LNC rates are coherently enhanced by a factor of~$2$;
    \item \textbf{Majorana-like}: the $\Delta M \tau \gg 2\pi$ limit where both LNV and LNC processes are present, with the same integrated rates.
\end{itemize}
Comparing these two limiting cases for the same benchmark models allows to assess the level of uncertainty introduced by the unknown $\Delta M$.

\section{Procedure}
\label{sec:procedure}

In order to reinterpret the limits from the ATLAS prompt search~\cite{Aad:2019kiz} (with extra details in the Ph.D.\@ thesis \cite{thiele_atlas_2019}) we have tried to reproduce the ATLAS analysis as accurately as possible.
Our signal is simulated using \textsc{MadGraph5\_aMC@NLO}~\cite{Alwall:2014hca} with the \textsc{HeavyN} model~\cite{Alva:2014gxa,Degrande:2016aje} (\cref{sec:signal}).
For the event selection (\cref{sec:event_selection}), we have implemented the ATLAS cut flow and obtained comparable efficiencies (\cref{sec:efficiencies}).
We take the total background counts from the ATLAS publication~\cite{Aad:2019kiz} (\cref{sec:background}).
Finally, in order to compute the limits (\cref{sec:limits}), we use the \CLs{} test statistics, along with a very simplified treatment of uncertainties.

\subsection{Event selection}
\label{sec:event_selection}

The prompt ATLAS analysis~\cite{Aad:2019kiz} considers the final states consisting of three isolated charged leptons (with electron or muon flavor) with no opposite-charge same-flavor lepton pairs (in order to limit the background from $Z$ decays), \ie{} only $e^{\pm} e^{\pm} \mu^{\mp}$ (electron channel) and $\mu^{\pm} \mu^{\pm} e^{\mp}$ (muon channel) are considered. 
It focuses on HNLs which are sufficiently short-lived that their decay vertex can be efficiently reconstructed using the standard ATLAS tracking algorithm. 
Since our reinterpretation will include a number of processes not included in the original ATLAS analysis\footnote{By ``process'' we mean a set of diagrams which have the same incoming and outgoing particles (distinguishing in particular all possible flavors of outgoing (anti-)neutrinos). Thus, \eg{}, \cref{fig:LNV_LNC} corresponds to two different \emph{processes} in our terminology, although they share the same visible final state.
The same notion is used in \texttt{MadGraph}.\label{fn:process}} and having different kinematics (\eg{} LNC processes, which are absent in the single-flavor mixing assumption), we cannot use the published ATLAS efficiencies and we have to compute them on our own.

As we will see, imposing the same cut flow allows reproducing the ATLAS efficiencies with sufficient accuracy for the purpose of this reinterpretation.
The list of cuts is shown in \cref{tab:cutflow}, and their order roughly follows that of ref.~\cite{thiele_atlas_2019}.
When different cuts were applied to the 2015 and 2016 datasets, we use the 2016 cuts, since the 2015 dataset is smaller than the 2016 one by about an order of magnitude.
\begin{enumerate}
\item We start by applying a cut on the distance of closest approach to the origin in the $r$--$z$ plane: $|\Delta z_0 \sin(\theta)| < \SI{0.5}{mm}$ for the leading lepton\footnote{In the electron channel, the leading lepton is defined as the electron with the highest \pT{}, and in the muon channel as the muon with the highest \pT{}.} and $|\Delta z_0 \sin(\theta)| < \SI{1}{mm}$ for the remaining ones.
\item Next, we apply the default transverse momentum and pseudorapidity requirements on the three changed leptons, \ie{} $\pT > \SI{4.5}{GeV}$ and $|\eta| \in [0, 1.37[ \cup ]1.52, 2.47[$ for all electrons and $\pT > \SI{4}{GeV}$ and $|\eta| < 2.5$ for all muons.
\item Then, we simulate the selection performed by the trigger by applying the relevant $\pT{}$ requirements, as found in ref.~\cite{Aad:2019kiz}, ch.~4.1, \S{}1. For the single-electron trigger used in the electron channel, we do not apply the ID requirements, since the ID efficiency is difficult to accurately estimate.
\item We then apply the trigger offline requirements on the two leading leptons: $\pT(e_{\text{lead}}) > \SI{27}{GeV}$ and $\pT(e_{\text{sublead}}) > \SI{10}{GeV}$ for the electron channel and $\pT(\mu_{\text{lead}}) > \SI{23}{GeV}$ and $\pT(\mu_{\text{sublead}}) > \SI{14}{GeV}$ for the muon channel.
\item Next, we require the tri-lepton invariant mass $M_{3l}$ to be in the interval $]40, 90[\,\si{GeV}$.
\item We then apply a weight to each lepton in order to simulate the efficiency of lepton isolation. We use the \pT{}-differential isolation efficiencies reported in ref.~\cite{Aad:2019tso} for electrons and ref.~\cite{Aad:2020gmm} for muons, using the ``loose'' working point in both cases.
\item For the electron channel only, a further cut is applied on the invariant mass of the $e^{\pm}e^{\pm}$ pair, $M(e,e) < \SI{78}{GeV}$, in order to veto the background from $Z \to e^+ e^-$ where one of the electron charges is misreconstructed.
\item The missing transverse energy is then restricted to $E_{\mathrm{T}}^{\text{miss}} < \SI{60}{GeV}$.
\end{enumerate}
Finally, the events passing the above cuts are binned in $M(l_{\text{sublead}}, l')$, which approximates the invariant mass of the HNL for small HNL masses (for which the leading lepton is usually the prompt lepton). The bins are $[0, 10[$, $[10, 20[$, $[20, 30[$, $[30, 40[$ and $[40, 50[\,\si{GeV}$.

Our cut flow is summarized in \cref{tab:cutflow}. One notable difference with the ATLAS paper is the absence of a $b$-jet veto in our analysis, which we omitted since $b$-jets appear in only $\mathcal{O}(1\%)$ of the signal events, therefore this cut would remove almost no signal \emph{at truth level}. For this reason, we have not generated $b$-jets in our final samples.
A further difference comes from the cuts related to the displacement of the leading lepton. ATLAS additionally imposes $|d_0/\sigma(d_0)| < 5$ (electron) or $< 3$ (muon), while we only impose the $|\Delta z_0 \sin(\theta)|$ cut and omit the $d_0$ cut, since we do not know $\sigma(d_0)$ well enough.\footnote{In principle, the data on $\sigma(d_0)$ is reported in ref.\ \cite{ATL-PHYS-PUB-2020-005} (fig.~4).
However, it only exits for muons and  is too coarse-grained to be exploitable in our analysis.}
This most likely does not affect the signal at truth level, since the leading lepton has a very small displacement in all relevant cases: for light HNLs, the leading lepton is almost always the prompt lepton from the $W$ decay, while heavier HNLs decay with a very short displacement due to their much shorter lifetime.
We also decided to omit the lepton identification (ID) requirements, whose efficiency is harder to model for electrons due to being significantly less smooth~\cite{Aad:2019tso} than the isolation one, in particular for the ``tight'' working point and at low \pT{}. For muons the ID efficiency is close enough to 1~\cite{Aad:2020gmm} that it can probably be safely neglected. Our attempt at implementing this cut only resulted in a significantly decreased accuracy for the efficiency estimates. A possible cause could be that the tabulated efficiencies have been computed using different sets of triggers and cuts and therefore cannot be transposed directly to the present analysis.

\begin{table}
    \centering
    \begin{tabular}{l|c|c}
        \toprule
        \# & Electron channel & Muon channel \\
        \midrule
        1 & \multicolumn{2}{c}{$|\Delta z_0 \sin(\theta)|(l) < \begin{cases} \SI{0.5}{mm} \text{ for } l = l_{\mathrm{lead}} \\ \SI{1}{mm} \text{ for } l \neq l_{\mathrm{lead}} \end{cases}$} \\
        \hline
        \multirow{2}{*}{2} & \multicolumn{2}{c}{$\pT(e) > \SI{4.5}{GeV} {}^{\dagger}$, $\pT(\mu) > \SI{4}{GeV}$} \\ & \multicolumn{2}{c}{$\abs{\eta(e)} \in [0, 1.37[ \cup ]1.52, 2.47[$, $\abs{\eta(\mu)} < 2.5$} \\
        \hline
        \multirow{3}{*}{3} & & $\pT(\mu_{\mathrm{lead}}) > \SI{22}{GeV}{}^{\dagger}$ \\
        & $\pT(e_{\mathrm{lead}}) > \SI{26}{GeV}{}^{\dagger}$ & \textsc{and} \\
        & & $\pT(\mu_{\mathrm{sublead}}) > \SI{8}{GeV}$ \\
        \hline
        \multirow{2}{*}{4} & $\pT(e_{\text{lead}}) > \SI{27}{GeV}$ & $\pT(\mu_{\text{lead}}) > \SI{23}{GeV}$ \\
        & $\pT(e_{\text{sublead}}) > \SI{10}{GeV}$ & $\pT(\mu_{\text{sublead}}) > \SI{14}{GeV}$ \\
        \hline
        5 & \multicolumn{2}{c}{$\SI{40}{GeV} < M(l,l,l') < \SI{90}{GeV}$} \\
        \hline
        6 & \multicolumn{2}{c}{``Loose'' lepton isolation} \\
        \hline
        7 & $Z$~veto: $M(e,e) < \SI{78}{GeV}$ & --- \\
        \hline
        8 & \multicolumn{2}{c}{$E_{\mathrm{T}}^{\text{miss}} < \SI{60}{GeV}$} \\
        \bottomrule
    \end{tabular}
    \caption{Our cut flow for the electron and muon channels. The ${}^{\dagger}$ indicates cuts which differ between 2015 and 2016 (the 2016 cuts were used in this analysis). Lepton identification and $|d_0/\sigma(d_0)|$ cuts have been omitted due to the complexity of their implementation.}
    \label{tab:cutflow}
\end{table}

\subsection{Signal}
\label{sec:signal}

In order to reinterpret the sensitivity of the ATLAS prompt HNL search for arbitrary combinations of HNL masses $M_N$ and ratios of mixing angles, we need to be able to compute the expected signal counts in each $M(l_{\text{sublead}},l')$ bin in each signal region, for any model parameters. We do so using a simple model, described below.

\subsubsection{MadGraph setup}
\label{sec:MG_setup}

\begin{table}
    \centering
    \begin{tabular}{|l|c|c|c|l|}
        \hline
        \multicolumn{5}{|c|}{Electron channel ($e^{\pm} e^{\pm} \mu^{\mp}$)} \\
        \hline
        Process & $\Delta L$ & $\alpha$ & $\beta$ & \textsc{MadGraph} process string \\
        \hline
        $W^+ \to e^+ (N \to \mu^- e^+ \nu_e)$ & $0$ & $e$ & $\mu$ & \texttt{p p > e+ n1, n1 > mu- e+ ve} \\
        $W^- \to e^- (N \to \mu^+ e^- \bar{\nu}_e)$ & $0$ & $e$ & $\mu$ & \texttt{p p > e- n1, n1 > mu+ e- ve\textasciitilde} \\
        $W^+ \to e^+ (N \to e^+ \mu^- \bar{\nu}_{\mu})$ & $-2$ & $e$ & $e$ & \texttt{p p > e+ n1, n1 > e+ mu- vm\textasciitilde} \\
        $W^- \to e^- (N \to e^- \mu^+ \nu_{\mu})$ & $+2$ & $e$ & $e$ & \texttt{p p > e- n1, n1 > e- mu+ vm} \\
        \hline
    \end{tabular}
    \caption{Signal processes contributing to the electron channel. Up to two additional hard jets have been included in the process string, but are omitted here for brevity.}
    \label{tab:signal_electron_channel}

    \vspace*{\floatsep}

    \begin{tabular}{|l|c|c|c|l|}
        \hline
        \multicolumn{5}{|c|}{Muon channel ($\mu^{\pm} \mu^{\pm} e^{\mp}$)} \\
        \hline
        Process & $\Delta L$ & $\alpha$ & $\beta$ & \textsc{MadGraph} process string \\
        \hline
        $W^+ \to \mu^+ (N \to e^- \mu^+ \nu_{\mu})$ & $0$ & $\mu$ & $e$ & \texttt{p p > mu+ n1, n1 > e- mu+ vm} \\
        $W^- \to \mu^- (N \to e^+ \mu^- \bar{\nu}_{\mu})$ & $0$ & $\mu$ & $e$ & \texttt{p p > mu- n1, n1 > e+ mu- vm\textasciitilde} \\
        $W^+ \to \mu^+ (N \to \mu^+ e^- \bar{\nu}_e)$ & $-2$ & $\mu$ & $\mu$ & \texttt{p p > mu+ n1, n1 > mu+ e- ve\textasciitilde} \\
        $W^- \to \mu^- (N \to \mu^- e^+ \nu_e)$ & $+2$ & $\mu$ & $\mu$ & \texttt{p p > mu- n1, n1 > mu- e+ ve} \\
        \hline
    \end{tabular}
    \caption{Signal processes contributing to the muon channel. Up to two additional hard jets have been included in the process string, but are omitted here for brevity.}
    \label{tab:signal_muon_channel}
\end{table}

The signal processes contributing to each channel are listed in \cref{tab:signal_electron_channel,tab:signal_muon_channel}.\footnote{In the ``Process'' column, we use a bar to indicate the chirality of the produced light neutrinos. Their Majorana nature does not play a role here.} For Majorana-like HNL pairs, all processes contribute, while for Dirac-like HNL pairs only those which conserve the total lepton number ($\Delta L = 0$) contribute (with a factor-of-2 enhancement {for the total cross section}).

For each process, we generate a Monte-Carlo sample which will be used to compute both the cross section and the efficiency. Each sample consists of $\sim 40000$ weighted events generated at leading order using \textsc{MadGraph5\_aMC@NLO v2.8.x} \cite{Alwall:2014hca} along with the \textsc{HeavyN} model \cite{Alva:2014gxa,Degrande:2016aje} (specifically, we use the \textsc{SM\_HeavyN\_CKM\_AllMasses\_LO} model\footnote{Note that HNLs are Majorana particles in this model. This is actually not a problem for simulating quasi-Dirac HNLs: all we need to do is suitably rescale or suppress the cross section of each process, as will be discussed in \cref{sec:quasi-dirac}.}), which includes the non-diagonal CKM matrix as well as the finite fermion masses.
The center of mass energy is set to $\sqrt{s} = \SI{13}{TeV}$ and the integrated luminosity to $\mathcal{L}_{\text{int}} = \SI{36.1}{fb^{-1}}$, in order to match the parameters of the 2019 prompt analysis. We generate the processes listed in the ``\textsc{MadGraph} process string'' column in \cref{tab:signal_electron_channel,tab:signal_muon_channel}, with up to two additional hard jets (excluding $b$-jets). \textsc{Pythia 8} is then used (through the \textsc{MadGraph} interface) to shower and hadronize the events. We use the event weights and the merged cross section reported by \textsc{Pythia}.

\subsubsection{Signal computation for arbitrary model parameters}
\label{sec:scaling}

In order to obtain the physical cross section, a number of model parameters need to be specified: the HNL mass $M_N$, its mixing angles\footnote{Since we are dealing with 2 HNLs far from the seesaw line, $\Theta_{\alpha 2} \cong \pm i \Theta_{\alpha 1}$~\cite{Shaposhnikov:2006nn,Kersten:2007vk}. We generate the Monte-Carlo samples for a single HNL with parameters $\Theta_{\alpha} \eqdef \Theta_{\alpha 1}$, such that $|\Theta_{\alpha}| = |\Theta_{\alpha 1}| \cong |\Theta_{\alpha 2}| = \frac 12 U_\alpha^2$, see \cref{U2:def}.} $|\Theta_e|$, $|\Theta_{\mu}|$ and $|\Theta_{\tau}|$ and its total decay width $\Gamma_N$.
Generating a new sample for every set of parameters would be computationally prohibitive. Fortunately, we can leverage the scaling properties of the cross section in order to exactly recompute it for each new set of mixing angles.
This is done as follows.

As a first step, we generate Monte-Carlo samples for all the processes listed in \cref{tab:signal_electron_channel,tab:signal_muon_channel}, for each HNL mass $M_N \in \{5,10,20,30,50\}\,\si{GeV}$ and using the reference parameters $|\Theta|_{\text{ref}} = 10^{-3}$ and $\Gamma_{\text{ref}} = 10^{-5}\,\si{GeV}$ as placeholders for the remaining model parameters.\footnote{These parameters allow for the successful numerical integration in the narrow width approximation.}
For each process~$P$, we only set the relevant mixing angle $|\Theta_{\alpha(P)}|$ and $|\Theta_{\beta(P)}|$ to $|\Theta|_{\text{ref}}$, where $\alpha(P)$ and $\beta(P)$ respectively correspond to the generations coupling to the HNL at production and decay, as listed in \cref{tab:signal_electron_channel,tab:signal_muon_channel}.

The key observation here is that the branching fraction of $W^+ \to l_{\alpha} N$ is proportional to $|\Theta_{\alpha}|^2$, while the branching fraction of $N \to l_{\beta} l_{\gamma} \nu_{\gamma}$ is proportional to $|\Theta_{\beta}|^2 / \Gamma_N$. Therefore, the cross section for a given process $P$ is proportional to $|\Theta_{\alpha(P)}|^2 |\Theta_{\beta(P)}|^2 / \Gamma_N$. Starting from the reference cross section $\sigma_P^{\text{ref}}$ obtained for the reference parameters, this allows to extrapolate the physical cross section to new parameters:
\begin{equation}
    \sigma_P(M_N,\Theta_e,\Theta_{\mu},\Theta_{\tau}) = \sigma_P^{\text{ref}} \times \frac{|\Theta_{\alpha(P)}|^2 |\Theta_{\beta(P)}|^2}{|\Theta|_{\text{ref}}^4} \times \frac{\Gamma_{\text{ref}}}{\Gamma_N(M_N,\Theta_e,\Theta_{\mu},\Theta_{\tau})}
    \label{eq:sigma_P}
\end{equation}
Since the total HNL width enters this formula, we need to be able to compute it for arbitrary parameters too. To this end we follow a similar approach. We notice that the partial width into a given decay channel~$D$ is proportional to $|\Theta_{\beta(D)}|^2$, where $\beta(D)$ denotes the flavor with which the HNL mixes when decaying. Summing over all decay channels and all three flavors, we can then express the total decay width as:
\begin{equation}
    \tau_N^{-1} = \Gamma_N(M_N,\Theta_e,\Theta_{\mu},\Theta_{\tau}) = \sum_{\beta = e,\mu,\tau} |\Theta_{\beta}|^2 \times \hat{\Gamma}_{\beta}(M_N)
    \label{eq:total_width}
\end{equation}
where $\hat{\Gamma}_{\beta}(M_N) = \Gamma_N(M_N,\delta_{\beta e},\delta_{\beta\mu},\delta_{\beta\tau})$ is the total decay width obtained by setting $\Theta_{\beta} = 1$ and the two other mixing angles to zero.
It can be easily computed with \textsc{MadGraph} by generating the \texttt{n1 > all all all} process.
This extrapolation method, which makes use of the scaling properties of the relevant branching fractions, has been successfully validated by explicitly computing the cross section for a few non-trivial benchmark points and comparing the results.
The contribution $N_P$ of a given process~$P$ to the total event count (before applying any selection) is then obtained by multiplying the relevant cross section by the integrated luminosity: $N_P = \mathcal{L}_{\text{int}} \times \sigma_P$.

\subsubsection{Signal computation for quasi-Dirac HNLs}
\label{sec:quasi-dirac}

Finally, since the signal samples have been computed for a single Majorana HNL, we need to apply a correction factor~$c_P$ to each cross section when considering a quasi-Dirac HNL pair. If this HNL pair is Majorana-like (\ie{} it has both LNC and LNV processes with equal rates), then all cross sections must be multiplied by~$2$, since there are two mass eigenstates whose event rates add incoherently. However, for a Dirac-like HNL pair (which only has LNC processes), the LNC cross sections must be multiplied by~$4$ due to the coherent enhancement discussed in \cref{sec:quasi_dirac_limit}, while the LNV ones should all be set to zero. Unlike in the case of a single Dirac fermion, no correction to the total HNL width needs to be applied. The correction factors are summarized in \cref{tab:quasi_dirac_correction}.

\begin{table}
    \centering
    \begin{tabular}{|l|c|c|c|}
        \hline
        \textbf{Nature} & $c_P,\ P \in\text{LNC}$ & $c_P,\ P\in\text{LNV}$ & $c_{\Gamma} = \Gamma_N / \Gamma_{\mathrm{Maj.}}$ \\
        \hline
        One Majorana HNL (reference) & $\mathbf{1}$ & $\mathbf{1}$ & $\mathbf{1}$ \\
        One Dirac HNL & $1$ & $0$ & $1/2$ \\
        Quasi-Dirac pair: \textbf{Majorana-like} & $2$ & $2$ & $1$ \\
        Quasi-Dirac pair: \textbf{Dirac-like}\footnotemark{} & $4$ & $0$ & $1$ \\
        \hline
    \end{tabular}
    \caption{Multiplicative coefficients $c_P$ to be applied to the cross section of each process~$P$, and $c_{\Gamma}$ to be applied to the total HNL width~$\Gamma_N$, depending on the HNL(s) nature and on whether the process is LNC or LNV.}
    \label{tab:quasi_dirac_correction}
\end{table}
\footnotetext{Note how a quasi-Dirac pair in the $\delta M=0$ limit is equivalent to a single Dirac HNL, up to a redefinition of the mixing angles $\Theta_{\alpha I}^{\text{quasi-Dirac}} = \Theta_{\alpha I}^{\text{Dirac}} / \sqrt{2}$.}

\subsection{Efficiencies}
\label{sec:efficiencies}

In order to obtain a sensitivity estimate, we must compute the expected signal count in every $M(l_{\text{sublead}},l)$ bin reported by the ATLAS collaboration.\footnote{We consider both signal regions (for the $e^{\pm}e^{\pm}\mu^{\mp}$ and $\mu^{\pm}\mu^{\pm}e^{\mp}$ signatures) simultaneously, so there are $10$~bins in total: $5$ in the electron channel and $5$ in the muon channel.} This is done by multiplying the true signal count by a signal efficiency. Since the relative contributions of the various diagrams --- which all have different kinematics and therefore different efficiencies --- depend on the model parameters, in general we expect the signal efficiency to depend on the mass~$M_N$, nature (Majorana-like or Dirac-like), lifetime~$\tau_N$ and all the mixing angles of the quasi-Dirac HNL pair. However, when considering a single process/diagram, the nature and mixing angles ``factor out'' such that the efficiency for this process depends only on the mass and lifetime of the HNL. We therefore need to compute one efficiency $\epsilon_{P,b}(M_N,\tau_N)$ for every process~$P$ and every bin~$b$. The total event count in bin~$b$ is then computed by summing over all the processes:
\begin{equation}
    N_b = \mathcal{L}_{\text{int}} \times \sum_P \epsilon_{P,b}(M_N,\tau_N) \times c_P \times \sigma_P(M_N,\Theta_e,\Theta_{\mu},\Theta_{\tau})
    \label{eq:event_count_in_bin}
\end{equation}
where $c_P$ is the correction factor applied to the cross section for quasi-Dirac HNLs.

For a given process~$P$ and bin~$b$, the efficiency $\epsilon_{P,b}(M_N, \tau_N)$ is computed by filtering the corresponding Monte-Carlo sample through the cut flow described in \cref{sec:event_selection,tab:cutflow}.
The binned efficiency is then:
\begin{equation}
    \epsilon_{P,b} = \frac{\sum \text{(weights of events after cuts, which end up in bin $b$)}}%
    {\sum \text{(weights of all events before cuts, from any bin)}}
\end{equation}
where the sums run over all events generated for the process~$P$ and the
events which fail to pass a given cut have their weight set to zero.\footnote{Some cuts (such as lepton ID and isolation cuts) are implemented by reweighting events using tabulated efficiencies.} 
Similarly, we can obtain the unbinned efficiency as:
\begin{equation}
    \epsilon_P = \frac{\sum \text{(all event weights after cuts)}}{\sum \text{(all event weights before cuts)}}.
\end{equation}

\begin{figure}
    \centering
    \includegraphics[width=\textwidth]{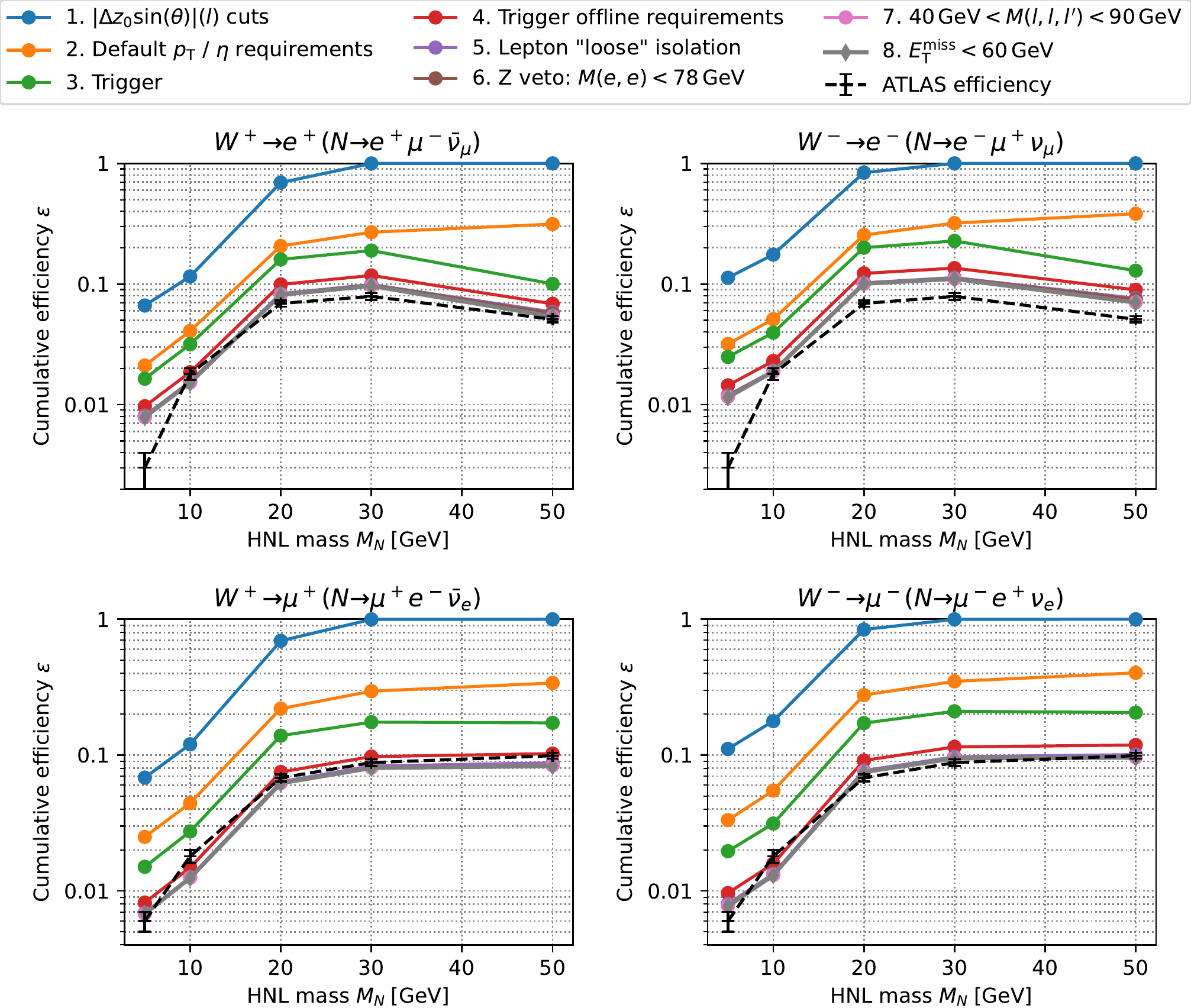}
    \caption{Cumulative unbinned signal efficiencies (for the total event count, \ie{} summed over all bins) after applying each cut listed in \cref{tab:cutflow}, computed for the benchmark points found in ref.~\cite{Aad:2019kiz}. The black dashed line denotes the total efficiencies reported in ref.~\cite{Aad:2019kiz}, table~2, and should be compared to the gray line with diamond markers (which corresponds to all cuts being applied). These efficiencies are for lepton number \emph{violating} (LNV) processes only, since these were the only relevant processes in the original prompt search.}
    \label{fig:efficiencies_lnv}
\end{figure}

\begin{figure}
    \centering
    \includegraphics[width=\textwidth]{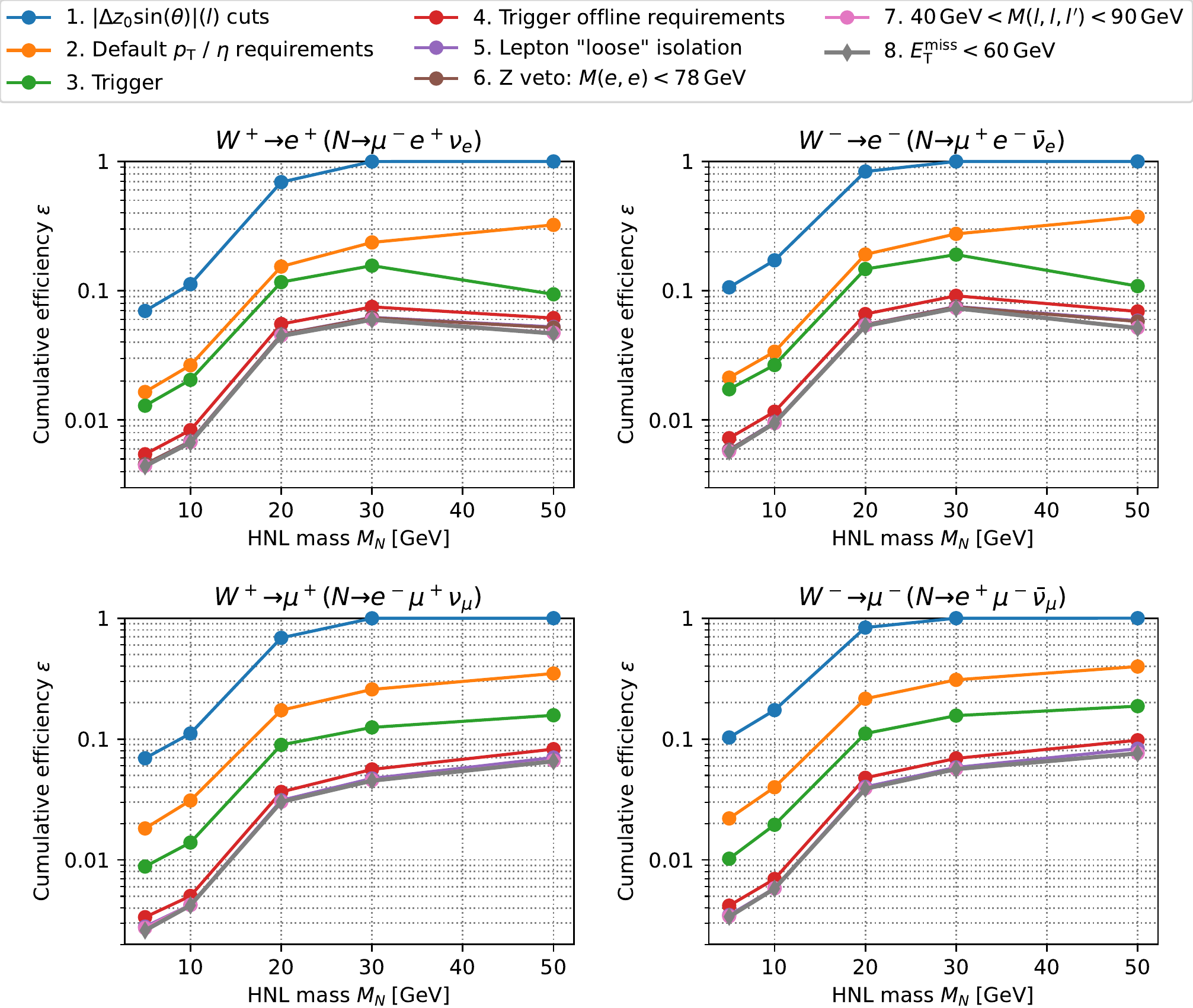}
    \caption{Cumulative unbinned signal efficiencies (for the total event count, \ie{} summed over all bins) after applying each cut listed in \cref{tab:cutflow}, computed for the benchmark points found in ref.~\cite{Aad:2019kiz}, for lepton number \emph{conserving} (LNC) processes. The gray line with diamond markers corresponds to the total efficiency.}
    \label{fig:efficiencies_lnc}
\end{figure}

The unbinned efficiencies for the four LNV processes are plotted in \cref{fig:efficiencies_lnv} along with the efficiencies reported by ATLAS in ref.~\cite{Aad:2019kiz}, while those for LNC processes are plotted in \cref{fig:efficiencies_lnc}. Since the efficiency of a process depends on both the HNL mass and its lifetime, we had to choose a set of benchmark points to produce \cref{fig:efficiencies_lnc,fig:efficiencies_lnv}. In order to be able to compare our efficiency calculation with the ATLAS efficiencies, we have chosen the same benchmarks as reported in ref.~\cite{Aad:2019kiz} and reproduced in \cref{tab:fabian_benchmarks}.
Our estimate is reasonably accurate for the muon channel, with a mean relative error\footnote{We define the relative error on the total efficiency $\epsilon$ as $\frac{|\epsilon_{\mathrm{ours}}-\epsilon_{\mathrm{ATLAS}}|}{\frac{1}{2} (\epsilon_{\mathrm{ours}}+\epsilon_{\mathrm{ATLAS}})}$.} of $18\%$ (maximum $48\%$), but less so for the electron channel, with a mean relative error of $38\%$ and a factor of $\sim4$ in the worst case (which corresponds to the lowest HNL mass hypothesis $M_N = \SI{5}{GeV}$).
The main difference between the two channels comes from the larger reliance on the electron ID (which we ignore) in the electron channel.
Indeed, the electron ID  is used for the single-electron trigger as well as for the ID cuts on both electrons; and contrary to the ``loose'' muon ID \cite[][fig.~12]{Aad:2020gmm} used for muons, its efficiency can be significantly smaller than~$1$ \cite[][fig.~17]{Aad:2019tso}. 
This omission could contribute to the worse agreement between signal efficiencies in the electron channel.
Another potential factor could be the large HNL displacement.
The displacement has not been taken into account when tabulating the isolation efficiencies (computed for $Z \to ll$ in refs.\ \cite{Aad:2019tso,Aad:2020gmm}). 
This would explain why the discrepancy is stronger for larger $c\tau_N\gamma_N$.
Comparing \cref{fig:efficiencies_lnc} with \cref{fig:efficiencies_lnv}, we also notice that \textbf{the efficiencies for LNC processes can be significantly smaller than for LNV processes}. This is mostly due to the different spin correlation patterns \cite{Tastet:2019nqj,Ruiz:2020cjx} for LNC vs.\@ LNV leading to different lepton spectra and to different geometrical acceptances of the lepton \pT{} and displacement cuts.

Since the original Monte-Carlo samples used for this analysis did not take spin correlations into account, 
and were generated under the single-flavor mixing hypothesis, 
the cut flow has been optimized under these assumptions.
In principle, this might lead to a sub-optimal cut selection when it is applied to more realistic models (which we eventually hope to \emph{observe}). For this reason, we would generally recommend performing the cut optimization using a set of signal samples which are representative of realistic models (instead of simplified benchmarks) and which have been generated using a Monte-Carlo event generator (such as \textsc{MadGraph}) which can model spin correlations.
However, in the present case, it seems that most cuts were chosen solely based on the minimal requirements imposed by the existing triggers, and therefore would not have been very different had the cut optimization been performed with more realistic models in mind.

\begin{table}
    \centering
    \begin{tabular}{|l|c|c|c|c|c|}
        \hline
        HNL mass $M_N$ & \SI{5}{GeV} & \SI{10}{GeV} & \SI{20}{GeV} & \SI{30}{GeV} & \SI{50}{GeV} \\
        \hline
        HNL lifetime $\tau_N$ & \SI{1}{mm} & \SI{1}{mm} & \SI{0.1}{mm} & \SI{0.01}{mm} & \SI{1}{\micro m} \\
        \hline
    \end{tabular}
    \caption{Benchmark points (taken from ref.~\cite{Aad:2019kiz}) used to plot the efficiencies in \cref{fig:efficiencies_lnv,fig:efficiencies_lnc}. Note that our calculations are more general, and work for any combination of $M_N$ and $\tau_N$.}
    \label{tab:fabian_benchmarks}
\end{table}

\begin{figure}
    \centering
    \includegraphics[width=0.65\textwidth]{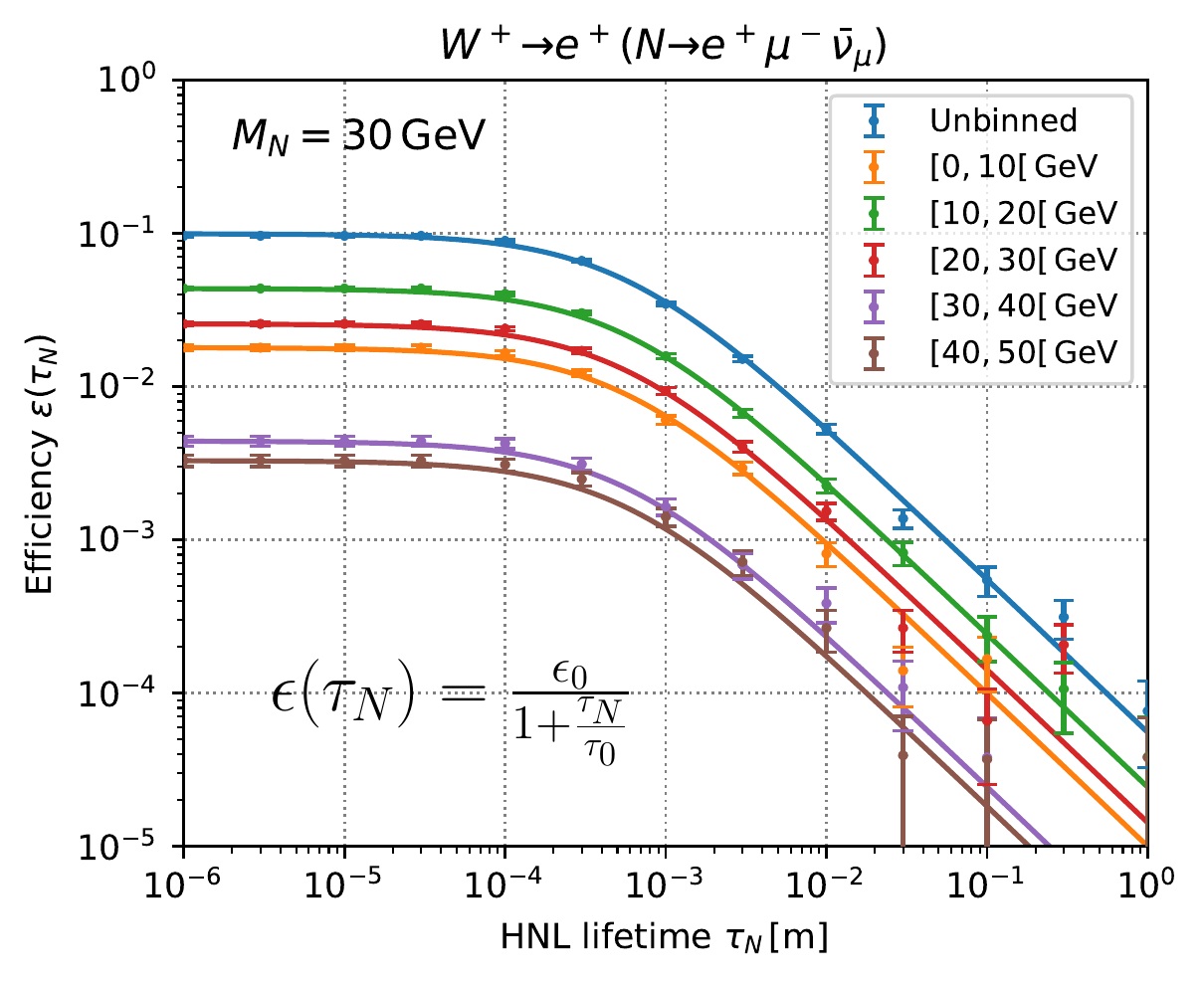}
    \caption{Binned and unbinned efficiencies as a function of the HNL lifetime $\tau_N$, for the process $W^+ \to e^+ (N \to e^+ \mu^- \bar{\nu}_{\mu})$ with $M_N = \SI{30}{GeV}$. The dots represent the efficiencies calculated explicitly, while the lines correspond to the fitted model. Error bars denote an \emph{estimate} of the statistical uncertainties from the finite size of the Monte-Carlo sample.}
    \label{fig:efficiency_vs_lifetime}
\end{figure}

Even using the extrapolation method described above and \cref{eq:event_count_in_bin}, one efficiency $\epsilon_{P,b}(M_N,\tau_N)$ must in principle still be computed for every process~$P$, bin~$b$, HNL mass $M_N$ and lifetime $\tau_N$. However, several simplifications exist. First, the efficiencies for the full set of $M(l_{\text{sublead}},l')$ bins (keeping the other parameters fixed) can be computed simultaneously, since the events only need to go through the cut flow once, before the binning is applied. More interestingly, it also turns out that the $\tau_N$ dependence can be quite accurately parametrized using a simple functional form $\epsilon(\tau_N)$. This functional form can be constrained by requiring the following asymptotic behavior:
\begin{itemize}
    \item $\epsilon(\tau_N) \to \epsilon_0$ (prompt efficiency) as $\tau_N \to 0$.
    \item $\epsilon(\tau_N) \propto \frac{1}{\tau_N}$ for sufficiently large $\tau_N$.
\end{itemize}
The ``simplest'' functional form satisfying these two conditions is:
\begin{equation}
    \epsilon(\tau_N) = \frac{\epsilon_0}{1 + \frac{\tau_N}{\tau_0}}
    \label{eq:displaced_efficiency_parametrization}
\end{equation}
with $\epsilon_0$ the prompt efficiency and $\tau_0$ the typical lifetime after which the efficiency starts to drop due to the HNL displacement. After fitting it to the efficiencies which have been explicitly computed for a number of lifetime points, this model can be used to extrapolate the efficiency to arbitrary HNL lifetimes. As an example, the model, along with the lifetime points used for the fit, are presented in \cref{fig:efficiency_vs_lifetime} for both the binned and unbinned efficiencies, for the $W^+ \to e^+ (N \to e^+ \mu^- \bar{\nu}_{\mu})$ process with a \SI{30}{GeV} HNL. The relative error between the data and the model is $\lesssim 10\%$ (on top of the statistical error). The efficiencies for other processes and mass points display a similar behavior.

Thanks to these simplifications, for each HNL mass~$M_N$ and process~$P$, the efficiencies need only be computed for $3$ or more lifetime points (we used $13$) in order to obtain the full lifetime dependence along with an error estimate. This amounts to $12$ or more Monte-Carlo samples per mass point for Dirac-like HNL pairs, and $24$ or more for Majorana-like HNL pairs.\footnote{Plus three samples for computing the HNL lifetime, but these only need to be run at parton level and therefore have a negligible computational cost.} Lifetime reweighting can additionally be used to simulate intermediate lifetimes without having to generate new samples. This makes the approach computationally tractable (although expensive) for experiments who would like to report their efficiencies in a benchmark-agnostic way, while still using their full detector simulation.

\subsection{Background}
\label{sec:background}

A number of Standard Model processes can mimic the signatures that we are looking for. This can happen if these processes have the same final state (irreducible background) or if they are misreconstructed as the same final state (reducible background) due to \emph{fake} leptons (\ie{} non-prompt leptons from jets or leptons from pileup). ATLAS has found the irreducible background to be subdominant~\cite{Aad:2019kiz}, and the main background components to be \emph{multi-fakes} (multiple fake leptons coming from $W$+jets or multiple jets) as well as $t\bar{t}$ with a fake lepton.

Each of these background sources comes with statistical uncertainties. The kinematic distribution of the multi-fake sample is estimated from data using a number of \emph{estimation} regions, then normalized by fitting a normalization factor $\mu_{\text{mf}}$ to the three control regions. Due to the finite sizes of the data samples, both of these steps introduce statistical errors into the multi-fake estimate, with potentially non-trivial correlations between the $M(l_{\text{sublead}},l')$ bin counts, which we are ultimately interested in. Similarly, the finite size of the $t\bar{t}$ Monte-Carlo sample and the finite event counts in the control regions used to estimate its normalization factor $\mu_{t\bar{t}}$ also introduce statistical errors into the $t\bar{t}$ estimate.

The detailed uncertainties (including correlations) of the individual background components are not listed in ref.~\cite{Aad:2019kiz}. 
Performing a detailed background analysis is out of the scope of the present paper. Instead, we have decided to use a simplified background model, which only takes into account the total background count in each bin, but is nonetheless capable of providing a good enough approximation of the sensitivity for the purpose of this reinterpretation.

To this end, the total background count in each channel and each $M(l_{\text{sublead}},l')$ bin, along with its uncertainty band, is digitized from figure~5 in ref.~\cite{Aad:2019kiz}.
Since the uncertainties on the individual components of the background are unfortunately not reported, implementing a statistical test necessarily requires some guessing on our side. After experimenting with several well-motivated background models and selecting the one which leads to the best approximation of the ATLAS limits, we have decided to model the uncertainty as being entirely caused by a single, Gaussian-constrained normalization factor $\mu_{\text{tot}}$. In other words, we assume that the background expectations in the various $M(l_{\text{sublead}},l')$ bins are maximally correlated. This is consistent with the observation that the statistical errors on the normalization factors $\mu_{\text{mf}}$ and $\mu_{t\bar{t}}$ are among the leading uncertainties. The accuracy of this simplified model will be explicitly tested in \cref{sec:limits}.

\subsection{Statistical limits}
\label{sec:limits}

Ref.~\cite{Aad:2019kiz} found a very good compatibility between the observed counts and the background-only hypothesis. 
They then proceeded with exclusion limits by testing the compatibility of the observed counts under the signal $+$ background hypotheses for five different benchmark points in the (mass, lifetime) space, each for two different mixing patterns: with electron or muon flavor.

In order to define the exclusion limit, ATLAS uses the \CLs{} test~\cite{Read:2002hq}.
For completeness, a quick reminder about the \CLs{} technique follows in \cref{sec:general_CLs}.
Knowledgeable users are welcome to skip it and go directly to \cref{sec:our_CLs}.
\subsubsection{\texorpdfstring{\CLs{} technique: a general reminder}{CLs technique: a general reminder}}
\label{sec:general_CLs}
The \CLs{} technique is based on the likelihood-ratio test statistics, more specifically on:
\begin{equation}
    t(x) \equiv 2 \ln\left(\frac{\mathcal{L}(x|H_{s+b})}{\mathcal{L}(x|H_b)}\right)
\end{equation}
where $\mathcal{L}$ denotes the likelihood, $x$ the data, $H_b$ the background-only hypothesis and $H_{s+b}$ a signal + background hypothesis. Larger values of $t$ indicate more signal-like data. The distribution of $t$ is estimated under each hypothesis through the use of pseudo-experiments~$X$: $p_b(t) = \mathcal{P}(t(X))$ for $X \sim H_b$ and $p_{s+b}(t) = \mathcal{P}(t(X))$ for $X \sim H_{s+b}$. Given an observation $x_{\text{obs}}$ and the corresponding value of the test statistics $t_{\text{obs}} = t(x_{\text{obs}})$, the $\mathrm{CL}_b$ and $\mathrm{CL}_{s+b}$ values are then computed as:
\begin{align}
    \mathrm{CL}_b &= \mathcal{P}\left(t(X) < t_{\text{obs}} | H_b\right) = \int_{-\infty}^{t_{\text{obs}}} \d{t} p_b(t) \\
    \mathrm{CL}_{s+b} &= \mathcal{P}\left(t(X) < t_{\text{obs}} | H_{s+b}\right) = \int_{-\infty}^{t_{\text{obs}}} \d{t} p_{s+b}(t)
\end{align}
In other words, $\mathrm{CL}_b$ and $\mathrm{CL}_{s+b}$ are the probabilities of obtaining a dataset that is more background-like than the observed one, respectively under the background and signal + background hypotheses. Both increase for increasingly signal-like $x_{\mathrm{obs}}$.
Finally, the value of the \CLs{} test statistics is given by the ratio:
\begin{equation}
    \CLs = \frac{\mathrm{CL}_{s+b}}{\mathrm{CL}_b} \in [0,1]
\end{equation}
and a given signal + background hypothesis $H_{s+b}$ is considered to be excluded if $\CLs < 0.05$. For any signal stronger than the $\CLs = 0.05$ limit, the probability of a type-I error (false exclusion) will always be less than $0.05$.
In order to complete the statistical analysis, the likelihood remains to be specified. 
We will proceed with this in the following section.

\subsubsection{\texorpdfstring{\CLs{} technique: implementation}{CLs technique: implementation}}
\label{sec:our_CLs}
The observables in question are the event counts in the two signal regions (for the electron and muon channels), each channel consisting of $5$ $M(l_{\text{sublead}},l')$ bins. 
Since we will be dealing with non-trivial combinations of mixing angles, we simultaneously include both channels in our likelihood. 
We thus end up with $10$ bin counts $\{x_i\}$, with $i=1\dots 5$ for the electron channel and $i=6\dots 10$ for the muon channel.
As discussed in \cref{sec:background}, we model the background as a set of expectation values $\{b_i\}$ for each bin~$i=1\dots 10$ (taken from the ATLAS paper) along with a Gaussian-constrained normalization factor $\mu_{\text{tot}}$ with standard deviation $\sigma_{\text{tot}} = \sum_i ( b_i^+ - b_i^-) / (2\sum_i b_i)$, where the ${}^-$ and ${}^+$ superscripts respectively denote the lower and upper uncertainty bands from the ATLAS plot (see \cref{tab:ATLAS_bg}).
The signal is modeled as a set of signal expectations $\{s_i\}$, $i=1\dots 10$, which we compute for each set of the model parameters $(M_N,\Theta_e,\Theta_{\mu},\Theta_{\tau})$ using the method described in \cref{sec:signal,sec:efficiencies}. 
Contrary to ATLAS, we do not use a signal strength parameter~$\mu$, since this would amount to rescaling the mixing angles without changing the lifetime, leading to inconsistent results.\footnote{In the prompt limit ($\tau_N \equiv 0$), the approach taken by ATLAS would work. However, HNLs in the lowest two mass bins ($5$ and \SI{10}{GeV}) have a small displacement, which can strongly affect the efficiency.} We neglect all uncertainties on the signal counts, which we have estimated to be at the sub-percent level.
The bin counts~$x_i$ are assumed to be Poisson distributed, with expectation values of respectively $\mu_{\text{tot}} b_i$ for the background-only hypothesis and $\mu_{\text{tot}} b_i + s_i$ for the signal + background hypothesis.
The full likelihood for the signal + background hypothesis is thus:
\begin{multline}
    \mathcal{L}(x|H_{s+b}) = \mathcal{P}(\mu_{\text{tot}} | \mathcal{N}(1, \sigma_{\text{tot}})) \times \prod_{i=1}^{10} \mathcal{P}(x_i | \text{Pois}(\mu_{\text{tot}} b_i + s_i)) \\ \text{where } \mu_{\text{tot}} = \frac{\sum_i (x_i - s_i)}{\sum_i b_i}
    \label{eq:likelihood}
\end{multline}
The likelihood for the background-only hypothesis $H_b$ is obtained by setting the signal $s_i$ to zero in \cref{eq:likelihood}.

\begin{figure}
    \centering
    \includegraphics[width=\textwidth]{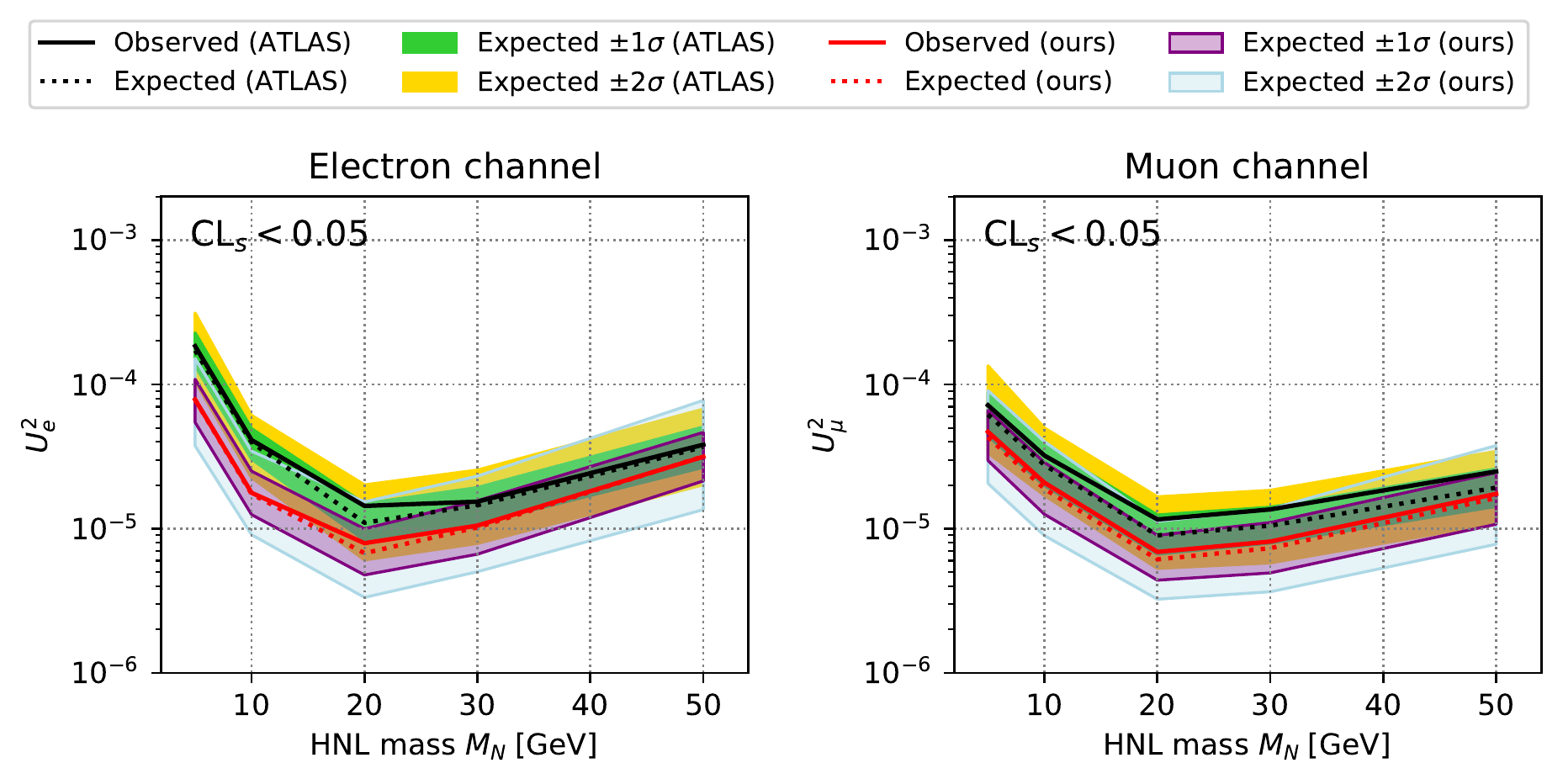}
    \caption{Comparison of the limits obtained using our simplified statistical model with the ones observed by ATLAS, using the exact same dataset (\ie{} event counts, total background and expected signal).}
    \label{fig:CLs_comparison}
\end{figure}

In order to validate our simplified statistical analysis, we can compare the limits that it produces to the limits obtained by ATLAS, when using the \emph{exact same counts} as ATLAS (extracted again from figure~5 in ref.~\cite{Aad:2019kiz}). In order to perform this comparison, a few changes need to be made. First, we need to reintroduce the signal strength parameter~$\mu$. Second, we need to consider both channels separately. After making these changes, we obtain the limits shown in \cref{fig:CLs_comparison}. The mean ratio between our limits and the ones from ATLAS is $0.64$, and the worst-case ratio is $0.42$. Although not fully satisfactory, this discrepancy should still be small enough to allow us to reliably compare limits which differ by an order of magnitude or more, as we will do in the next section. This is especially true when the reinterpreted limits are all computed using the same method. 
\begin{table}[!t]
    \centering
    \begin{tabular}{p{3em}|p{3em}|p{3em}cp{3em}|p{3em}|p{3em}c}
        \toprule
        \multicolumn{4}{c}{$ee\mu$} & \multicolumn{4}{c}{$\mu\mu e$} \\
        \cmidrule(lr){1-4}
        \cmidrule(lr){5-8}
        \multicolumn{3}{c}{Expected background} & Observed count & \multicolumn{3}{c}{Expected background} & Observed\\
        $b_i$ & $b_i^-$ & $b_i^+$ & $x_i$ & $b_i$ & $b_i^-$ & $b_i^+$ & $x_i$\\
        \cmidrule(lr){1-3}
        \cmidrule(lr){4-4}
        \cmidrule(lr){5-7}
        \cmidrule(lr){8-8}
        19.0 & 14.9 & 23.1 &  19 &  21.3 & 17.5 & 25.1 &  23 \\
        18.0 & 14.4 & 21.7 &  20 &  13.8 & 10.6 & 17.0 &  15 \\
        21.0 & 17.4 & 24.7 &  19 &  18.7 & 15.1 & 22.3 &  20 \\
        13.6 & 10.9 & 16.2 &  15 &  13.3 & 10.6 & 16.1 &  14 \\
        6.1 &  4.2 &  7.8 &   5 &  13.1 & 10.2 & 15.9 &  13 \\
        \bottomrule
    \end{tabular}
    \caption{Background in 5 invariant mass bins (rows) for the searches in $e^\pm e^\pm \mu^\mp$ and $\mu^\pm \mu^\pm e^\mp$ channels correspondingly. 
    The values have been digitized from Figure~5 in~\protect\cite{Aad:2019kiz}.
    Only the total background expectation (without the individual contributions) is shown.}
    \label{tab:ATLAS_bg}
\end{table}

\section{Results}
\label{sec:results}

\begin{figure}
    \begin{minipage}[c]{0.48\textwidth}
        \centering
        \includegraphics[width=\textwidth]{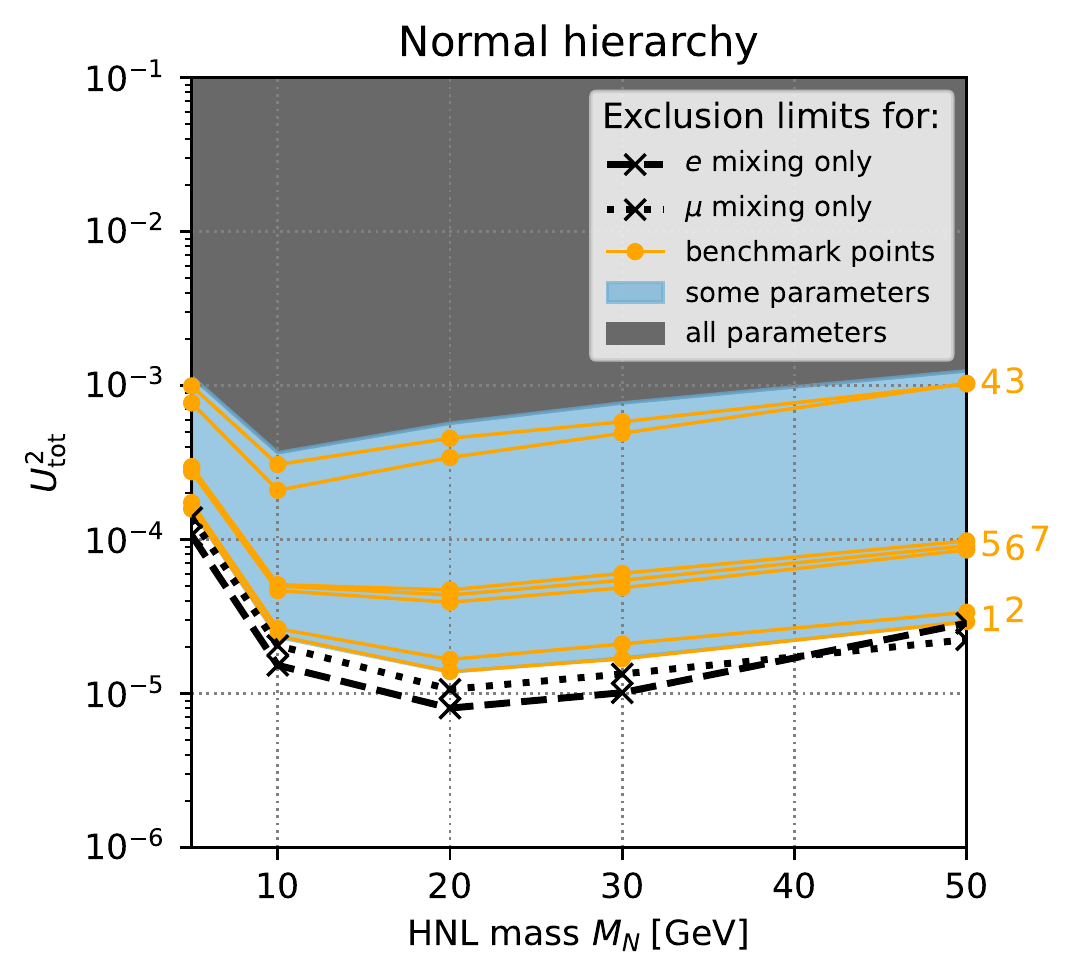}
    \end{minipage}
    \hfill
    \begin{minipage}[c]{0.48\textwidth}
        \centering
        \includegraphics[width=\textwidth]{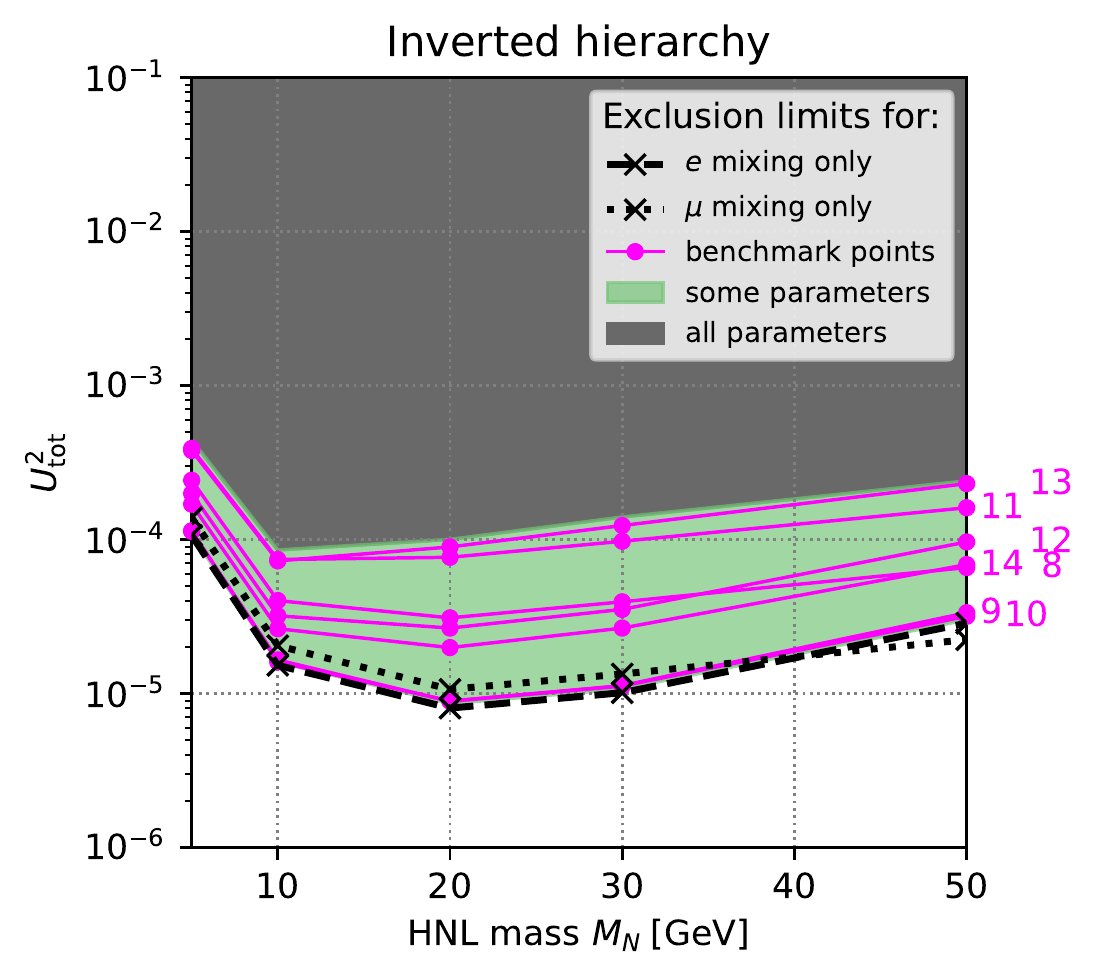}
    \end{minipage}
    \caption{Original (black lines) and reinterpreted (colored lines) $95\%$ exclusion limits on the total mixing angle $U_{\text{tot}}^2 = \sum_{\alpha=e,\mu,\tau} \sum_{I=1,2} |\Theta_{\alpha I}|^2$ for a \textbf{Majorana-like} HNL pair for the normal (left) and inverted (right) mass orderings. 
    The black lines are limits obtained under the single-flavor assumption, while the solid colored lines denote those obtained for the benchmark points defined in \cref{fig:ternary_plot_with_benchmarks}. 
    When scanning over all ratios of mixing angles allowed by neutrino oscillation data, the exclusion limits span the blue (green) shaded regions.
    Correspondingly, the gray filled area is excluded at $\mathrm{CL} > 95\%$ for all possible ratios of mixing angles, and thus constitutes an exclusion limit independent of the specific choice of mixing angles, valid as long as we consider the two HNL model explaining neutrino oscillations.}
    \label{fig:recast_limits_total_U2_Majorana}
\end{figure}

\begin{figure}
    \begin{minipage}[c]{0.48\textwidth}
        \centering
        \includegraphics[width=\textwidth]{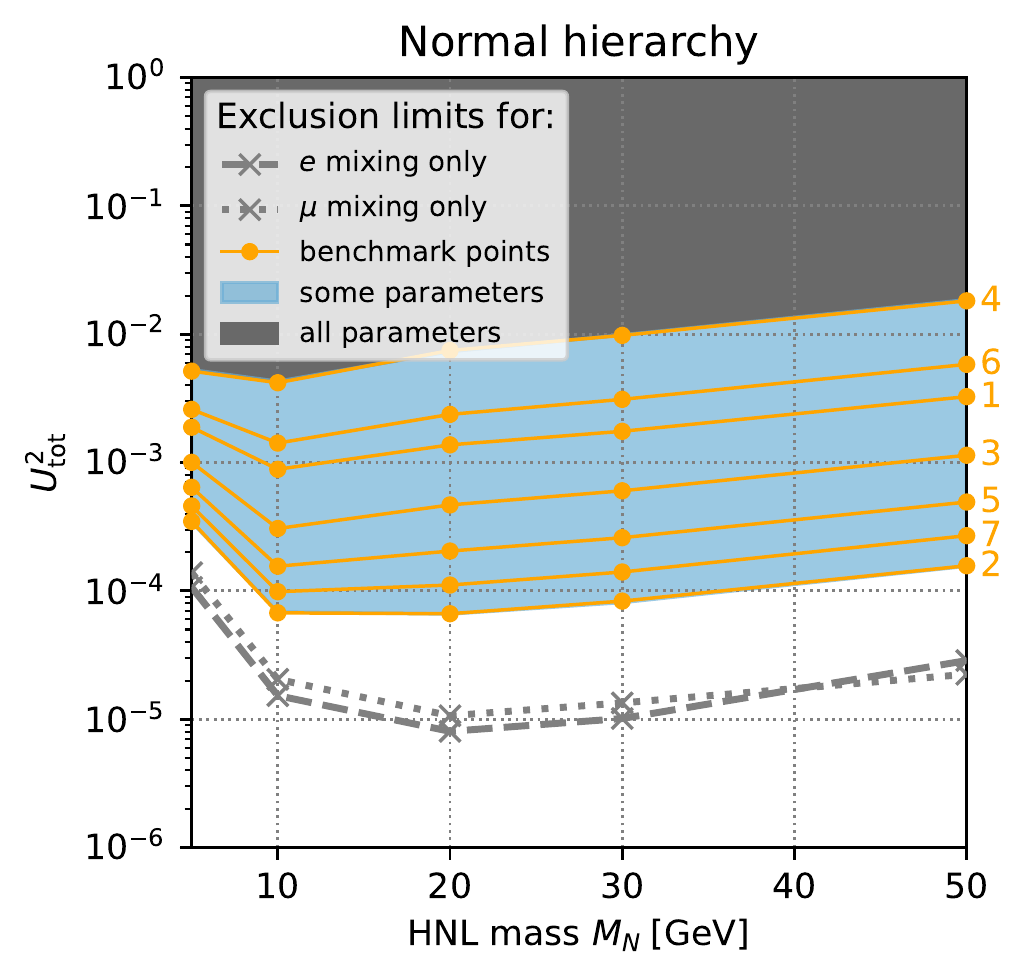}
    \end{minipage}
    \hfill
    \begin{minipage}[c]{0.48\textwidth}
        \centering
        \includegraphics[width=\textwidth]{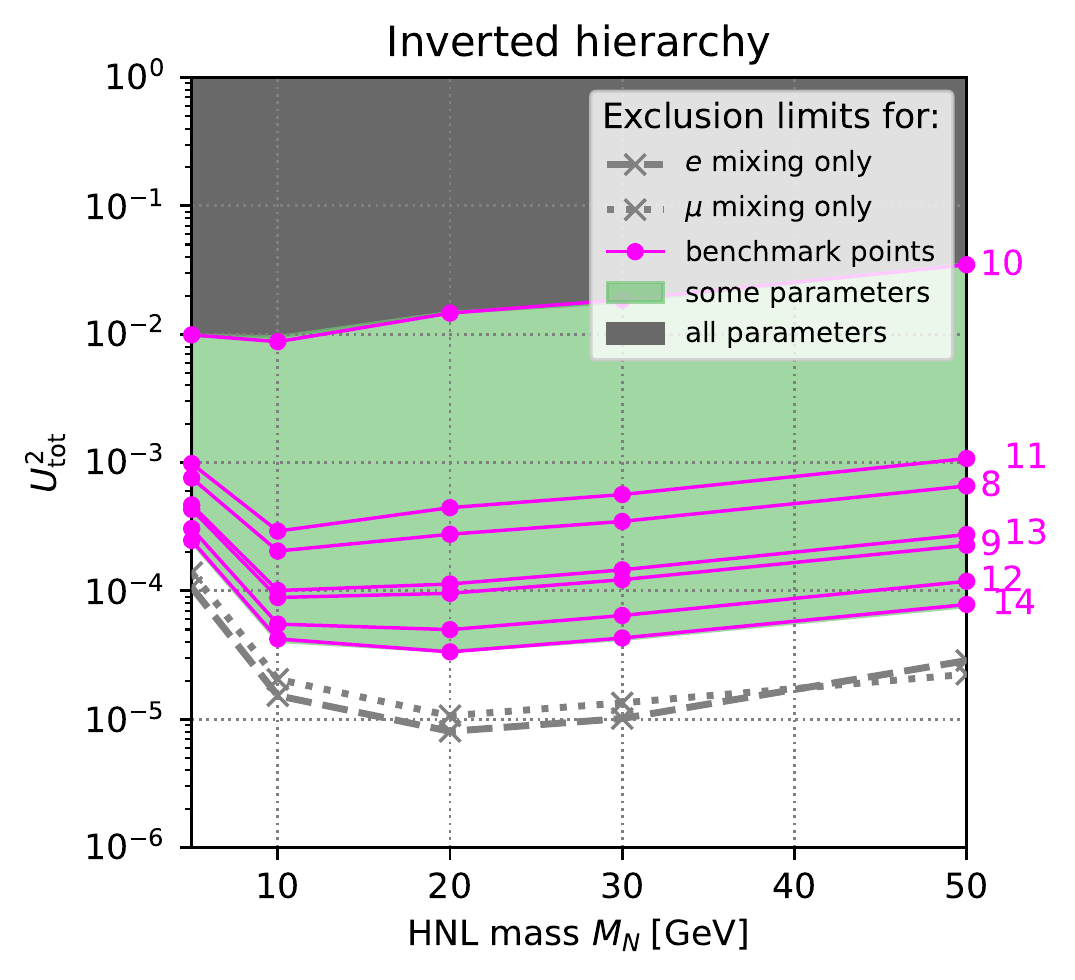}
    \end{minipage}
    \caption{Same as \cref{fig:recast_limits_total_U2_Majorana}, but for a \textbf{Dirac-like} HNL pair. The single-flavor mixing limits are grayed out because this search has \emph{no sensitivity} to the Dirac-like case under this assumption; instead the limits for the Majorana-like case are given for comparison.}
    \label{fig:recast_limits_total_U2_Dirac}
\end{figure}

Below we present our results --- the exclusion limits for the model with two HNLs.
We calculate exclusions for each of the benchmark points defined in \cref{fig:ternary_plot_with_benchmarks}.
Benchmarks are chosen in such as way as to represent both typical and extreme ratios of the mixing angles $U_e^2:U_{\mu}^2:U_{\tau}^2$.
As each benchmark fixes the mixing pattern, our results are most compactly expressed as exclusion limits for the total mixing angle $U_{\mathrm{tot}}^2 = U_e^2 + U_{\mu}^2 + U_{\tau}^2$ (\cref{eq:U2Imomega}). \Cref{fig:recast_limits_total_U2_Majorana,fig:recast_limits_total_U2_Dirac} present our results for the Majorana- and Dirac-like cases respectively. 
The limits for the flavor mixing angles $U_{\alpha}^2$ are presented in \cref{fig:recast_limits_Majorana_NH,fig:recast_limits_Majorana_IH,fig:recast_limits_Dirac_NH,fig:recast_limits_Dirac_IH}.
All these limits are the observed exclusion limits, and all of them (including the single-flavor limits) have been derived using the same statistical method,\footnote{In this way all the limits in \cref{fig:recast_limits_total_U2_Majorana} to \cref{fig:recast_limits_Dirac_IH} are obtained by means of the same statistical procedure and under the same assumptions about the systematic uncertainties. Therefore, although they might slightly deviate from the actual ATLAS limits from ref.~\cite{Aad:2019kiz}, they should be comparable among themselves. For comparison of our limits with those derived by ATLAS, see \cref{fig:CLs_comparison}.} which we described in \ref{sec:limits}.

The legend for these plots is as follows. The thick dashed and dotted lines in each plot represent the exclusion limits obtained under the assumption of a Majorana-like HNL pair mixing with a single flavor (respectively the electron and muon flavor). Up to a factor of~2, this corresponds to the scenario considered by ATLAS in the current prompt search~\cite{Aad:2019kiz}.
These limits are grayed out in the plots for the Dirac-like pair in order to emphasize that the search has \emph{no sensitivity} to the Dirac-like case for the single-flavor mixing. 
The solid colored lines denote the exclusion limits obtained for the various benchmark points defined in \cref{fig:ternary_plot_with_benchmarks}. The benchmarks can be identified using the numbers in the right margin.
The colored, filled area represents the set of possible (``benchmark-dependent'')\footnote{The limits that we call ``benchmark dependent'' are valid for a specific ratio of mixing angles, while the ones we call ``benchmark independent'' have been obtained by marginalizing over all the combinations of mixing angles allowed by the neutrino oscillation data. The latter still rely on the general properties of the model: the number of HNLs, the neutrino mass ordering and whether the HNLs behave as a Dirac-like or Majorana-like particle. As such, they are still \emph{model} dependent.} limits spanned by all the combinations of mixing angles allowed by the \nufit{} neutrino data (at $95\%\,\mathrm{CL}$).\footnote{The confidence limit assumes the specified mass ordering and does not take into account ``priors'' on the mass orderings.}
In other words, it shows the dependence of the exclusion limits on the specific combination of mixing angles, within the constraints from neutrino oscillation data (which are represented by the similarly-colored area in \cref{fig:ternary_plot_with_benchmarks}). Finally, the gray filled area denotes the set of mixing angles which are excluded at the $95\%$ level for all the allowed ratios of mixing angles. It thus represents the most conservative (benchmark-independent) limit that can be obtained for a given model. No choice of mixing angles that is in agreement with neutrino oscillation data (within the 2~HNL seesaw model\footnote{This benchmark-independent limit would be much weaker for three HNLs, and non-existent for four or more HNLs, due to relaxed constraints from neutrino data.}) can produce a limit within the gray filled region.

\subsection{Majorana-like HNL pair}
\label{sec:results_majorana}

\begin{figure}
    \includegraphics[width=0.986\textwidth]{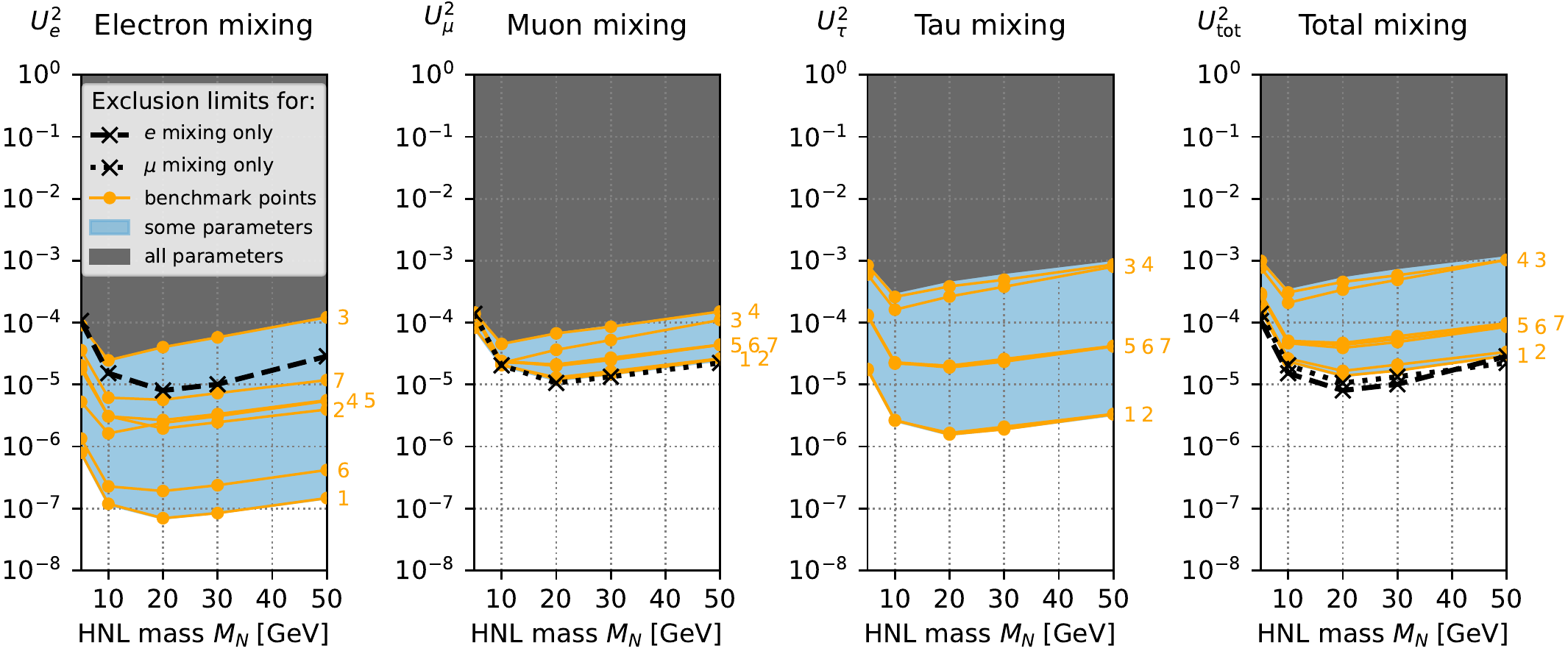}
    \caption{Original and reinterpreted exclusion limits (at $> 95\%$ CL) on the individual mixing angles $U_{\alpha}^2 = \sum_{I=1,2} |\Theta_{\alpha I}|^2 = x_{\alpha} U_{\mathrm{tot}}^2$ and the total mixing angle $U_{\mathrm{tot}}^2 = \sum_{\alpha=e,\mu,\tau} U_{\alpha}^2$ for a \textbf{Majorana-like} HNL pair and for the \textbf{normal hierarchy}. The legend is the same as in \cref{fig:recast_limits_total_U2_Majorana} and the rightmost figure coincides with the left panel in \cref{fig:recast_limits_total_U2_Majorana}.}
    \label{fig:recast_limits_Majorana_NH}
\end{figure}

\begin{figure}
    \includegraphics[width=1.000\textwidth]{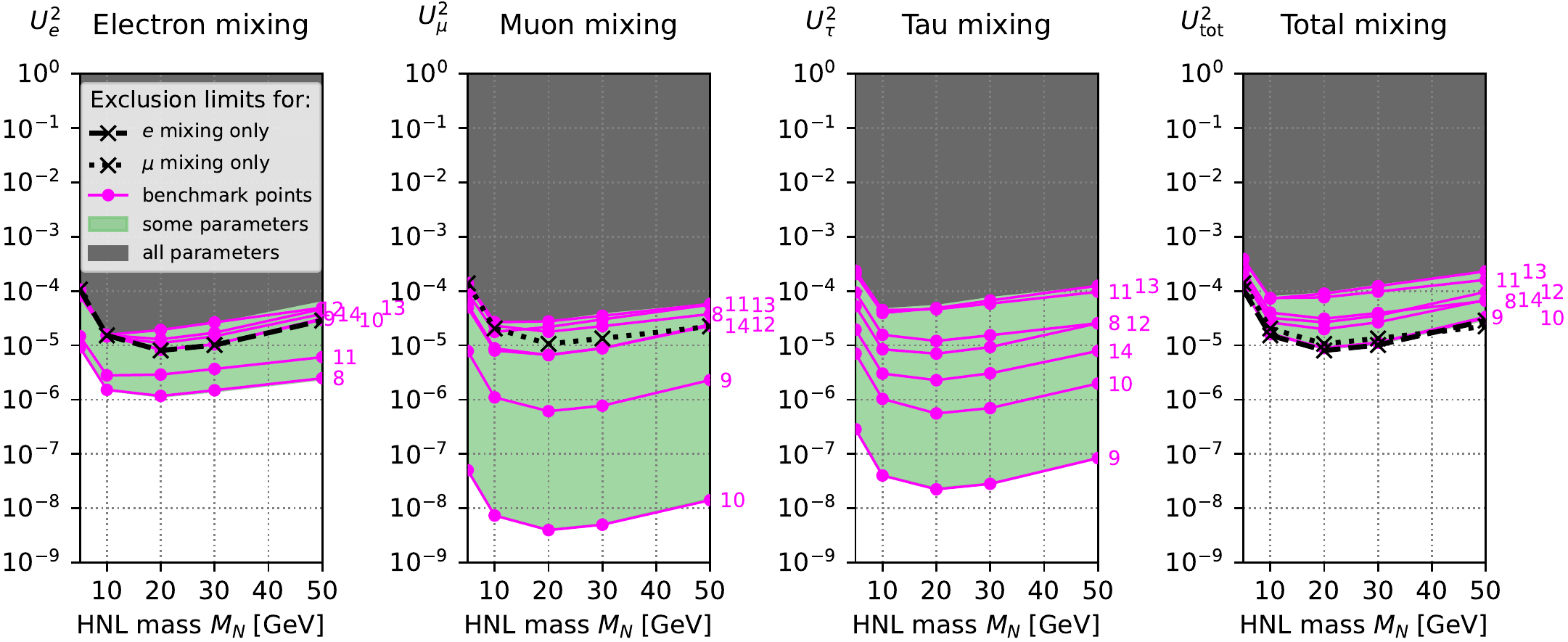}
    \caption{Same as \cref{fig:recast_limits_Majorana_NH}, for a \textbf{Majorana-like} HNL pair and the \textbf{inverted hierarchy}.}
    \label{fig:recast_limits_Majorana_IH}
\end{figure}

\begin{figure}
    \includegraphics[width=0.964\textwidth]{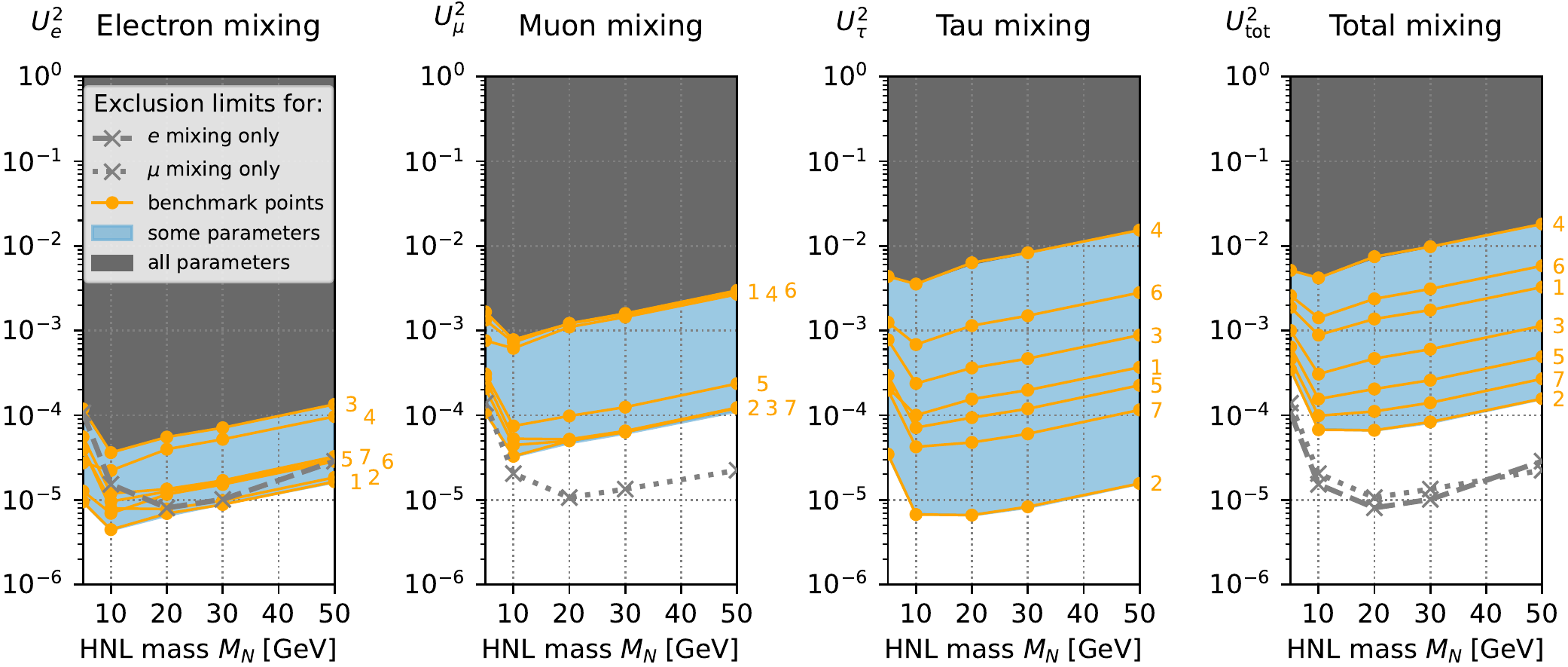}
    \caption{Same as \cref{fig:recast_limits_Majorana_NH}, for a \textbf{Dirac-like} HNL pair and the \textbf{normal hierarchy}. The legend is the same as in \cref{fig:recast_limits_total_U2_Dirac}.}
    \label{fig:recast_limits_Dirac_NH}
\end{figure}

\begin{figure}
    \includegraphics[width=0.983\textwidth]{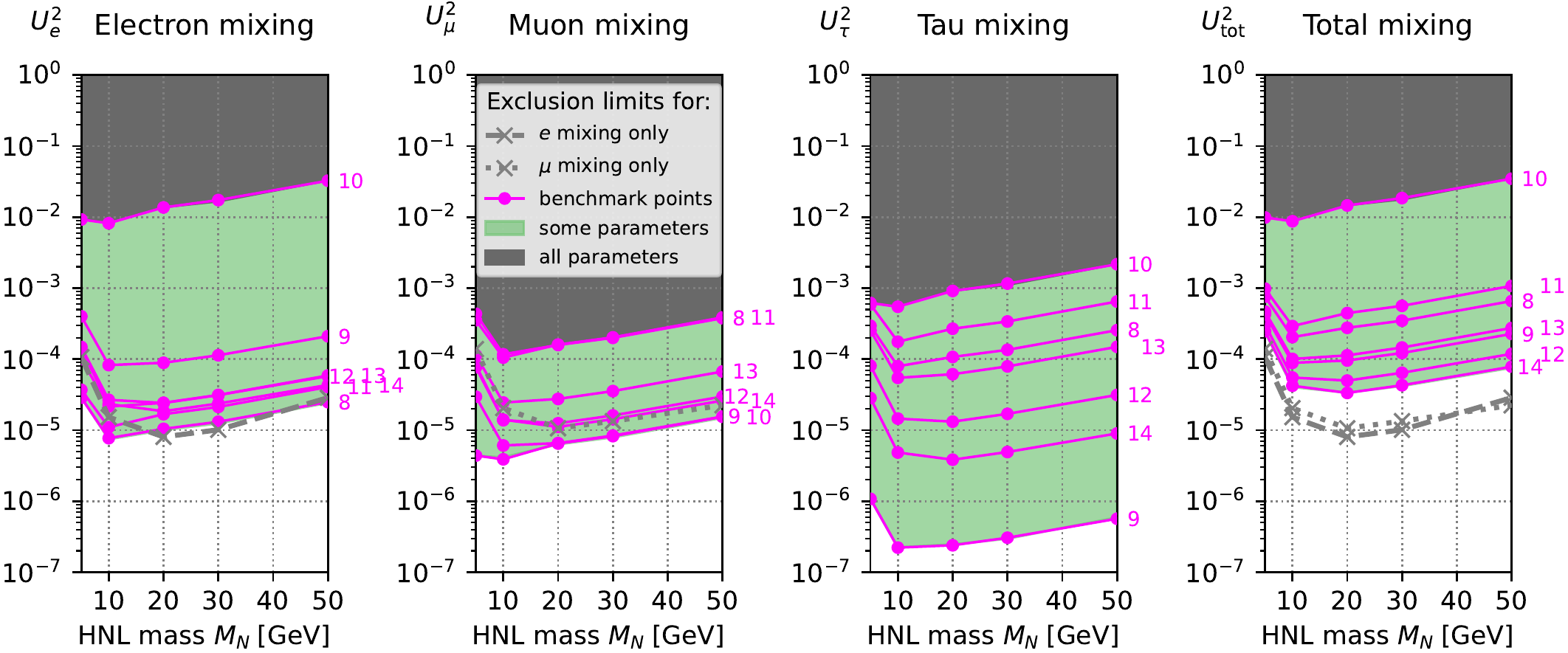}
    \caption{Same as \cref{fig:recast_limits_Dirac_NH}, for a \textbf{Dirac-like} HNL pair and the \textbf{inverted hierarchy}.}
    \label{fig:recast_limits_Dirac_IH}
\end{figure}

Let us first consider the case of a Majorana-like HNL pair, which is closer to the ``single Majorana HNL'' model considered by ATLAS and many other experiments. The relevant limits are shown in \cref{fig:recast_limits_total_U2_Majorana,fig:recast_limits_Majorana_NH,fig:recast_limits_Majorana_IH}. 
Apart from a trivial factor of two due to the two nearly degenerate mass eigenstates, the main difference with ATLAS is that in a realistic seesaw model the HNLs must mix with all three flavors at the same time. 
Looking at the total mixing angle in \cref{fig:recast_limits_total_U2_Majorana}, we immediately notice that the limits on $U_{\mathrm{tot}}^2$ are weaker than the single-flavor mixing limits for all our benchmarks, sometimes by more than an order of magnitude. The pattern is obvious for the normal hierarchy (but also visible for the inverted one): the benchmark points which have the strongest tau fraction $x_{\tau} = U_{\tau}^2/U_{\mathrm{tot}}^2$ also have the worst sensitivity. This was already observed in ref.~\cite{Abada:2018sfh}, and it is the manifestation of a well-known phenomenon: the introduction of new decay channels (here mediated by the tau mixing) reduces the branching fraction of the HNLs into the search channels. This has an important consequence: exclusion limits derived for $U_{\alpha}^2$ under the single-flavor assumption \emph{do not translate directly} into limits on $U_{\alpha}^2$ in a model where HNLs mix with multiple flavors.\footnote{Similarly, such limits do not apply if the HNLs have new interactions.} Instead, such limits \emph{must always be recast!}


When we look at the exclusion limits obtained for the individual mixing angles (in \cref{fig:recast_limits_Majorana_NH} for the normal hierarchy and \cref{fig:recast_limits_Majorana_IH} for the inverted hierarchy), we observe that for some benchmarks the exclusion limits on individual mixing angles can sometimes be \emph{much stronger} than the single-flavor limits. This actually reflects a rather trivial fact: if $U_\alpha^2 \ll U_{\beta}^2$ and the ratio $U_\alpha^2 : U_\beta^2$ is fixed, setting a limit on $U_\beta^2$ automatically sets a much stronger limit on $U_\alpha^2$ (\eg{} the limit set on $U_e^2$ for benchmark~$10$ in the IH indirectly sets a limit on $U_{\mu}^2$, which is enhanced by the ratio of the two mixing angles, in this case $U_{\mu}^2/U_e^2 \sim 1/2000$). 
In the same way we obtain an indirect limit (filled gray region) on the \emph{tau} mixing angle, which was not directly probed by this search. This simply reflects the fact that no valid combination of mixing angles which passes the constraints set by ATLAS in both the electron and muon channels, can have a mixing angle $U_{\tau}^2$ with tau above this limit. Although the fact that introducing new constraints (such as fixing the ratio of mixing angles) can increase the sensitivity is not unexpected, it may still be useful when one considers \emph{specific} sets of model parameters. This situation is not so far-fetched, since this is what happens when performing a scan over the parameter space in order to \eg{} combine constraints from multiple sources, which may be complementary if they probe different combinations of mixing angles. For instance, we expect that future experimental results (such as excluding one neutrino mass hierarchy, or observing/setting limits on neutrinoless double-beta decay) will introduce additional constraints on the possible combinations of mixing angles, thus leading to a more predictive model. These potential use cases once again support the reinterpretation of exclusion limits.

\subsection{Dirac-like HNL pair}
\label{sec:results_dirac}

Let us now turn our attention to the case of a Dirac-like HNL pair. Unlike in the Majorana-like case, there is \emph{no} observable lepton number violation in this case, since the HNLs do not have enough time to oscillate among themselves. Its phenomenology thus significantly differs from the one of a single Majorana HNL, usually considered by experiments. In particular, the only lepton number conserving contributions to the experimental signatures considered in ref.\ \cite{Aad:2019kiz} come from processes in which the HNL mixes with different flavors during its production and decay (due to the veto of opposite-charge same-flavor trilepton events). This search has therefore \emph{no sensitivity} to Dirac-like HNLs mixing with a single flavor!

By reinterpreting the limits (obtained for one Majorana HNL) within a realistic seesaw model (which \emph{requires} HNLs to mix with all three flavors), we are nonetheless able to set some exclusion limits for this model. These limits are presented in \cref{fig:recast_limits_total_U2_Dirac,fig:recast_limits_Dirac_NH,fig:recast_limits_Dirac_IH}. The legend is the same as for the Majorana-like HNL pair, except for the single-flavor mixing limits which are grayed out in order to emphasize that they were computed for a different model (Majorana-like HNLs) and are only present here for comparison purpose. Looking at our benchmark points, we immediately notice that the limits for the total mixing angle (\cref{fig:recast_limits_total_U2_Dirac}) are always weaker than the corresponding Majorana-like/single-flavor limits, sometimes by more than three orders of magnitude. The weakest limits are obtained when one of $U_e^2$ or $U_{\mu}^2$ is suppressed compared to the other, which is unsurprising given that this approximates the single-flavor mixing case, to which the search has no sensitivity. Looking at the colored, filled area, we also observe a wider possible range of limits (with variations by more than two orders of magnitude) compared to the Majorana-like case, depending of the specific ratio of mixing angle chosen. This reflects the fact that the limits now depend mainly on two mixing angles instead of just one, which enhances the benchmark dependence. Finally, similarly to the Majorana-like case, we observe that we can obtain strong benchmark-dependent limits on the individual mixing angles (see \cref{fig:recast_limits_Dirac_NH,fig:recast_limits_Dirac_IH}), as well as some benchmark-independent limits (for this specific seesaw model with a Dirac-like HNL pair; see the gray filled area). The latter are significantly weaker (by up to two orders of magnitude) than for a Majorana-like HNL pair, due to the larger variation among benchmarks.

We can summarize the case of Dirac-like HNLs by emphasizing how, despite the absence of sensitivity to the single-flavor mixing case, we nonetheless managed to obtain both benchmark-dependent and benchmark-independent (but still model-dependent) exclusion limits by reinterpreting the ATLAS results within a realistic seesaw model featuring a Dirac-like HNL pair. Since the relevant processes now depend on the product of two different mixing angles, limits for Dirac-like HNLs show a stronger dependence on their ratio than limits for Majorana-like HNLs, resulting in weaker benchmark-independent exclusion limits (filled gray area) for this model. Yet, the reinterpretation allowed us to obtain a limit on all three mixing angles (as well as their sum), where there was previously none (from this search).

\section{Conclusion \& outlook}
\label{sec:conclusion}

\subsection{Reinterpretation}

Heavy neutral leptons (HNLs) are promising candidates for explaining neutrino masses and oscillations. Within the seesaw model, their mass scale is not predicted by neutrino masses. 
Experiments searching for HNLs typically report null results in the form of exclusion limits on the mixing angle with one of the lepton flavors.
We emphasize that these constraints are neither model nor benchmark independent. 
Rather they correspond to limits obtained \emph{within a specific model where one HNL mixes with a single flavor}. 
As discussed in \cref{sec:seesaw}, these simplified models are incompatible with the observed neutrino masses and mixing pattern.
One may then wonder if the exclusion limits reported within these models remain valid when considering more realistic and theoretically motivated models of HNLs.
In this work, we have performed a \emph{reinterpretation} of the latest ATLAS prompt search for heavy neutral leptons~\cite{Aad:2019kiz} within one of the simplest \emph{realistic} models: a low-scale seesaw mechanism with two quasi-degenerate HNLs. 
At least two HNLs are required in order to be compatible with neutrino oscillation data, and the combination of their mixing angles is constrained by the seesaw relation. In particular, for two HNLs, no mixing angle can be zero.

Our aim was to study to which extent the exclusion limits on the HNL mixing angles are model or benchmark dependent and by how much they change when considering our more realistic model.
To this end, we have implemented a simplified version of the analysis employed by ATLAS in ref.~\cite{Aad:2019kiz}.
This reinterpretation was described in details in \cref{sec:procedure}.

Furthermore, as discussed in \cref{sec:quasi_dirac_limit}, the two HNLs must form a ``quasi-Dirac'' pair (\ie{} be nearly degenerate, with a specific mixing pattern) for sufficiently large mixing angles (which may be accessible at current experiments) to be viable.
Depending on the specific value of the mass splitting as well as the length scale over which the HNLs are observed, this quasi-Dirac pair may behave either as a Majorana-like or a Dirac-like particle, due to quantum interference between the two mass eigenstates.
Only Majorana-like HNL pairs feature lepton number violating decays, and the different spin correlation patterns for LNC and LNV decay chains lead to different signal efficiencies for Majorana-\@ and Dirac-like HNLs. Moreover, due to the veto applied by ATLAS on opposite-charge same-flavor lepton pairs (in their prompt HNL search), different diagrams, which depend on different combinations of mixing angles, contribute to the signal regions for Majorana-\@ and Dirac-like HNLs. In particular, the only diagrams contributing to the signal in the case of Dirac-like HNLs involve two different mixing angles, such that there is no sensitivity at all under the single-flavor mixing assumption! In order to handle both the Majorana-\@ and Dirac-like cases, we have performed the reinterpretation for each of them separately. The results were respectively presented in \cref{sec:results_majorana,sec:results_dirac}.

\paragraph{For Majorana-like HNL pairs, we have observed that:}
\begin{itemize}
    \item The exclusion limit on the total mixing angle $U_{\mathrm{tot}}^2$ is always weaker (sometimes by more that one order of magnitude) in realistic models than for single-flavor mixing. This is essentially caused by the opening of new decay channels (hence reducing the other branching fractions) which do not contribute to the search signature.
    \item Fixing the ratio of the mixing angles can result in (sometimes significantly) stronger \emph{indirect} constraints on some of the mixing angles. This can be useful when performing scans over the model parameters.
    \item Assuming the two-HNLs seesaw model and marginalizing over the ratio of mixing angles while keeping the HNL mass fixed, we can obtain limits on the individual mixing angles (including the tau mixing angle, which was not probed directly by this search) which do not depend on their ratio.
\end{itemize}

\paragraph{For Dirac-like HNL pairs, we have observed that:}
\begin{itemize}
    \item Contrary to the single-flavor mixing where the signal was identically zero, in our realistic model no single mixing angle can ever be zero, which ensures that we can always set an indirect (model-dependent) limit.
    \item The limits on the total mixing angle are, however, always weaker (by up to three orders of magnitude) than in the Majorana-like, single-flavor case.
    \item The weakest limits are obtained when one of $U_e^2$ or $U_{\mu}^2$ is suppressed compared to the other. This is expected, since these mixing patterns approximate the single-flavor case.
    \item Compared to the Majorana-like case, the dependence of the limits on the specific benchmark is stronger. This is likely caused by the fact that the product of two different mixing angles enters the cross section as a factor (instead of a single mixing angle) thus enhancing the parametric dependence.
    \item Similarly to the Majorana-like case, we can also set strong benchmark-dependent limits on the individual mixing angles by fixing their ratio. However, the corresponding marginalized/benchmark-independent limits are significantly weaker (by up to two orders of magnitude) due to the increased benchmark dependence.
\end{itemize}

Our results show that the reinterpretation of the exclusion limits is a necessary step in order to test HNL models which differ from those directly probed by an experiment.
In particular, if one interprets the reported limits on some parameter in a given model as exclusion limits on the same parameter in a \textit{different} model, they risk wrongly excluding part of the parameter space within the latter.
This of course does not affect the validity of the limits set by the experiment for the ``one-HNL, single-flavor'' benchmarks; it just means that one should be cautious when investigating models other than those two initial benchmarks.

When assuming specific choices of model parameters (as in parameter scans), stronger constraints can often be derived for the individual mixing angles. In the case of two HNLs, benchmark-independent constraints can also be derived by marginalizing over all the combinations of mixing angles allowed by neutrino data. For three or more HNLs, we expect most of the above results to remain valid, with the notable exception of the marginalized limits, which become much weaker or even non-existent due to the significantly weaker constraints from neutrino data~\cite{Abada:2018oly,Chrzaszcz:2019inj}.

For experimental results to be useful for constraining a wide range of model and parameters, it is therefore desirable to cast them into a form which allows them to be easily reinterpreted, bearing in mind that the main ``drivers'' for such interpretations --- theorists --- are typically unfamiliar with the inner workings of the experiment.
Below we outline a concrete proposal for reporting these results in the case of heavy neutral leptons, that would allow for an easy reinterpretation of the exclusion limits.

\subsection{Wish-list for a painless reinterpretation of future experimental results}

The LHC collaborations typically conduct searches in terms of simplified models.
Theorists, on the other hand, investigate models which address some of the shortcomings of the SM. Those are typically more complicated, and it is therefore necessary to reinterpret the search results in order to test them.
In order to facilitate this reinterpretation, one would greatly benefit from the following data being reported alongside the analysis (see also the recommendations in refs.\ \cite{Abdallah:2020pec,HistFactory,PHYSTAT032021,Bierlich:2019rhm}):
\begin{compactitem}[$\bullet$]
    \item The \textit{observed bin counts}.
    \item The various \textit{efficiencies} needed to evaluate the signal using the method described in \cref{sec:signal,sec:efficiencies}, \ie{}:
    \begin{compactitem}[--]
        \item The prompt efficiency $\epsilon_{P,b}^0$ for every \emph{process}~$P$ (as defined above, see \vref{fn:process}) and every bin~$b$ in all signal regions. 
        In simple cases there is a one-to-one correspondence between a \textit{Feynman diagram} and a \textit{process}~$P$, as in the charge-current decays considered in this paper. Ideally, all possible processes contributing to the search signature should be included.
        In the present case this would mean:
        \begin{inparaenum}[\it (a)]
        \item single and mixed flavor processes;
        \item LNV and LNC processes;
        \item processes mediated by charge currents, neutral currents and by their interference.
        \end{inparaenum}
        \item If the parametrization in \cref{eq:displaced_efficiency_parametrization} (or a modification thereof) allows reproducing the actual efficiency \emph{even approximately}, report the relevant parameters such as the lifetime cutoff~$\tau_0$ in our case.
    \end{compactitem}
    This slightly differs from the recommendations of the LHC Reinterpretation Forum~\cite{Abdallah:2020pec}, which advocates for releasing the object-level efficiencies in order to enable more general reinterpretations. Since the scope of the present reinterpretation is restricted to HNL models, those are not needed, and instead the signal can be more easily and accurately estimated using the simplified signal extrapolation method presented in \cref{sec:signal,sec:efficiencies}.\footnote{Note that our proposed reinterpretation method and the one advocated by the LHC Reinterpretation Forum are complementary: they are located in different parts of the ``spectrum'' of reinterpretations presented in ref.~\cite{PHYSTAT032021}.}
    However, we agree with their recommendation to (among many other things) break down the efficiencies for each signal region (or bin~$b$), each topology or final state, and each particle lifetime $\tau$. This directly corresponds to our $\epsilon_{P,b}(\tau)$ if we include neutrinos in the final state $P$ (which we called ``process'' to avoid confusion with the \emph{visible} final state often used by experiments).
    As an example of how to report these per-process, per-bin efficiencies, \cref{sec:signal_efficiencies} describes the JSON files containing the efficiencies computed using our simplified cut flow. A similar layout could be used to report the actual signal efficiencies from the experiment.
    \item For the background it is important to release \textit{the likelihood function}. This can be either:
    \begin{compactitem}[--]
        \item The ``full'' likelihood, including every background component and nuisance parameter used in the analysis (to the extent that this is possible). This can be done using tools such as \texttt{HistFactory}~\cite{HistFactory,ATL-PHYS-PUB-2019-029} or \texttt{pyhf}~\cite{Heinrich:2021gyp}.
        \item A simplified likelihood, containing only the dominant background components and nuisance parameters (see \eg{} ref.~\cite{CMS-LH:2017} or the \texttt{simplify}~\cite{simplify-hep} package).
        \item The covariance matrix of the background~\cite{CMS-LH:2017}, for all the signal bins, across \emph{all signal regions} (since they need to be fitted together when considering non-trivial models with \eg{} both electron and muon mixing).
    \end{compactitem}
    This is in line with the recommendations from the LHC Reinterpretation Forum~\cite{Abdallah:2020pec}.
    Finally, to ensure that the reported likelihood is accurate enough for performing a reinterpretation, it is important to validate it, \eg{} by comparing the resulting limits with those obtained using the full analysis.
\end{compactitem}
To go further and to recast the analysis to a different class of models, which include Feynman diagrams not initially considered, one needs to be able to re-implement the cut flow, rather than use the efficiencies themselves.
This requires knowing the efficiency maps for non-trivial cuts such as ID and isolation (as a function of both $\pT$ and $\eta$). These maps should be \emph{conditional} on the cuts which appear before them in the cut flow, \ie{} they should be computed after applying the cuts appearing before them. This is in line with the recommendation of the LHC Reinterpretation Forum to report \emph{analysis-specific} efficiencies~\cite{Abdallah:2020pec}.

\acknowledgments

JLT and OR are thankful to  C.~Appelt, W.~Buttinger, M.~Danninger, H.~Lacker, D.~Trischuk, M.~Wielers, S.~Xella for many useful discussion regarding various aspects of the ATLAS HNL analysis.
JLT would like to thank F.~Thiele for the many explanations regarding his Ph.D.\ thesis.
JLT is grateful to G.~Lanfranchi and M.~Shaposhnikov for their constructive comments on his own Ph.D.\ thesis. IT is thankful to J.~Klari{\'c} and K.~Urqu\'{i}a for helpful comments on the manuscript.
This project has received funding from the European Research Council (ERC) under the European Union's Horizon 2020 research and innovation programme (GA 336581, 694896) and from the Carlsberg foundation. The work of IT and JLT has been supported by ERC-AdG-2015 grant 694896 and by the Swiss National Science Foundation Excellence grant 200020B\underline{ }182864.

\appendix

\section{Ancillary files}
\label{sec:data_files}

In order to simplify the interpretation of experimental results within realistic HNL models, we are including a number of data files along with the present publication. They can be used to generate the relevant signal samples, or to implement the extrapolation method presented in \cref{sec:signal}.
These files can be found in the companion Zenodo record~\cite{zenodo}.

\subsection{Card files for the Monte-Carlo event generation}

The \texttt{/attachments/card\_files} folder contains the \textsc{MadGraph} card files (ending in \texttt{.dat}) and scripts (ending in \texttt{.txt}) for generating the signal samples used in this analysis, as well as for computing the total HNL width.
Due to the OSSF veto, only processes with no opposite-charge same-flavor lepton pairs have been included. Additional relevant processes can easily be added by modifying the \texttt{generate} and \texttt{add process} lines in the \texttt{*.txt} files.
All samples (except the ones used to compute the HNL width, which are generated at parton level) are generated at leading order, include up to two hard jets, and are showered and hadronized using \textsc{Pythia 8}. This is essential for obtaining a realistic $W$ spectrum. The shower parameters could probably benefit from further tuning, and further improvements in the $W$ spectrum accuracy are expected at NLO (using a suitable model).
To allow computing the signal efficiencies, all cuts have been disabled in the run card (with the exception of the maximum $|\eta_{\mathrm{jet}}|$ which needs to be set to $5$ for correct matching).

\subsection{Signal cross sections}

The cross sections for the various processes considered in this analysis, as well as the total HNL width (both computed using \textsc{MadGraph} as described in \cref{sec:signal}), are provided as JSON files in the \texttt{/attachments/cross\_sections} folder.

The file \texttt{total\_hnl\_width.json} contains the total HNL width $\hat{\Gamma}_{\alpha}(M_N)$ (expressed in \si{GeV}), computed for the $5$ mass points used in this analysis, and under the assumption of unit mixing with a single flavor $\alpha$, for each flavor. The total HNL width can then be computed for any combinations of mixing angles using \cref{eq:total_width}.
The file is organized as two nested dictionaries, with the first key denoting the HNL mass $M_N$, and the second one the flavor $\alpha$ for which the total width $\hat{\Gamma}_{\alpha}(M_N)$ has been computed for a unit mixing angle $|\Theta_{\alpha}|^2 = 1$ (with \texttt{Wtot\_e} for $\alpha=e$, \texttt{Wtot\_mu} for $\mu$ and \texttt{Wtot\_tau} for $\tau$).

The file \texttt{cross\_sections.json} contains the reference cross sections $\sigma_P^{\mathrm{ref}}$ (in \si{pb}) for all the processes~$P$ considered in this analysis, expressed for $|\Theta|_{\mathrm{ref}}^2 = 1$ and $\Gamma_{\mathrm{ref}} = 10^{-5}\,\si{GeV}$.
The file is organized as two nested dictionaries, with the first key denoting the HNL mass $M_N$ and the second the process~$P$. The correspondence between the key and the physical process can be found in \cref{tab:process_keys}.

\begin{table}
    \centering
    \begin{tabular}{l|l}
        \toprule
        Key & Process \\
        \midrule
        \texttt{lnc\_e+mu-e+} & $W^+ \to e^+ (N \to \mu^- e^+ \nu_e)$ \\
        \texttt{lnc\_e-mu+e-} & $W^- \to e^- (N \to \mu^+ e^- \bar{\nu}_e)$ \\
        \texttt{lnc\_mu+e-mu+} & $W^+ \to \mu^+ (N \to e^- \mu^+ \nu_{\mu})$ \\
        \texttt{lnc\_mu-e+mu-} & $W^- \to \mu^- (N \to e^+ \mu^- \bar{\nu}_{\mu})$ \\
        \texttt{lnv\_e+e+mu-} & $W^+ \to e^+ (N \to e^+ \mu^- \bar{\nu}_{\mu})$ \\
        \texttt{lnv\_e-e-mu+} & $W^- \to e^- (N \to e^- \mu^+ \nu_{\mu})$ \\
        \texttt{lnv\_mu+mu+e-} & $W^+ \to \mu^+ (N \to \mu^+ e^- \bar{\nu}_e)$ \\
        \texttt{lnv\_mu-mu-e+} & $W^- \to \mu^- (N \to \mu^- e^+ \nu_e)$ \\
        \bottomrule
    \end{tabular}
    \caption{Correspondence between the process key and the actual process.}
    \label{tab:process_keys}
\end{table}

\subsection{Signal efficiencies}
\label{sec:signal_efficiencies}

The efficiencies resulting from the event selection described in \cref{sec:event_selection}, as well as their parametrization according to \cref{eq:displaced_efficiency_parametrization} (as discussed in \cref{sec:efficiencies}) can respectively be found in the files \texttt{efficiencies.json} and \texttt{fitted\_efficiencies.json} in the \texttt{/attachments/\allowbreak{}efficiencies} folder.

The file \texttt{efficiencies.json} is organized as follows. The data is located in a triply nested dictionary under the \texttt{data} key: the first level corresponds to the HNL mass hypothesis $M_N$, the second to the process key (\cf{} \cref{tab:process_keys}) and the third to the $M(l_{\mathrm{sublead}},l')$ bin for which the efficiency is computed. The values of the bottom-most dictionary are lists containing the efficiencies for a number of HNL lifetimes, as listed in meters in \texttt{levels/lifetime}.

Finally, the file \texttt{fitted\_efficiencies.json} is also organized as a triply nested dictionary, with the first level corresponding to the HNL mass $M_N$, the second to the process key, and where the third level denotes the fit parameter from \cref{eq:displaced_efficiency_parametrization}. \texttt{tau0} is for $\tau_0$, \texttt{epsilon0\_total} for $\epsilon_0$ (the unbinned prompt efficiency), and \texttt{epsilon0\_binned} is a list containing the prompt efficiencies $\epsilon_{0,b}$ for the five $M(l_{\mathrm{sublead}},l')$ bins~$b$ (in the same order as in \texttt{efficiencies.json}). The layout described here (or a similar one) can be used by experiments to report their signal efficiencies in a way that ensures that theorists will be able to compute the expected signal for arbitrary choices of mixing angles.


\bibliographystyle{JHEP}
\bibliography{bibliography}

\end{document}